\documentclass[aps,prl,twocolumn,groupedaddress,longbibliography]{revtex4-2}

\usepackage{outlines}
\usepackage{lineno}
\usepackage{dsfont}
\usepackage{enumitem}
\usepackage{braket}
\usepackage{amsmath}
\usepackage{mathtools}
\usepackage{amsfonts,amssymb,amsbsy}
\usepackage{graphicx}
\usepackage{MnSymbol}
\usepackage{xcolor}
\usepackage{hyperref}

\makeatletter
\let\saved@includegraphics\includegraphics
\AtBeginDocument{\let\includegraphics\saved@includegraphics}
\renewenvironment*{figure}{\@float{figure}}{\end@float}
\makeatother

\hypersetup{
    colorlinks=true,
    linkcolor=blue,
    filecolor=magenta,      
    urlcolor=red,
}

\begin{document}

\title{Demonstration of the trapped-ion quantum CCD computer architecture}

\author{J. M. Pino}
\email{juan.pino@honeywell.com}
\author{J. M. Dreiling}
\author{C. Figgatt}
\author{J. P. Gaebler}
\author{S. A. Moses}
\author{M. S. Allman}
\author{C. H. Baldwin}
\author{M. Foss-Feig}
\author{D. Hayes}
\author{K. Mayer}
\author{C. Ryan-Anderson}
\author{B. Neyenhuis}

\affiliation{Honeywell Quantum Solutions, 303 S. Technology Ct., Broomfield, Colorado 80021, USA}

\maketitle

The trapped-ion QCCD (quantum charge-coupled device) proposal \cite{Wineland98,Kielpinski02} lays out a blueprint for a universal quantum computer. The proposal calls for a trap device to confine multiple arrays of ions, referred to as ion crystals. Quantum interactions between ions in a single crystal occur via electric- or magnetic-field induced spin-motion coupling. Interactions between ions initially in different crystals are achieved by using precisely controlled electric fields to split the crystals, transport the ions to new locations, and then combine them to form new crystals where the desired interactions are applied. High-fidelity physical transport is the key to two advantages of the QCCD architecture: high-fidelity interactions between distant qubits and low crosstalk between gates. High-fidelity interactions between distant qubits are enabled by transport that is fast relative to the long coherence times of the qubit, chosen as atomic clock (magnetic-field insensitive) states, as well as the insensitivity of the ion's internal degrees of freedom to the electric fields used for transport. Low crosstalk is achieved by the physical separation between qubits in different locations, thereby preserving the low error rates demonstrated in small trapped-ion experiments~\cite{Gaebler16,Ballance16,Christensen19}. However, engineering a machine capable of executing these operations across multiple interaction zones with low error introduces a host of difficulties that have slowed progress in scaling this architecture to larger qubit numbers. Using a Honeywell cryogenic surface trap, we report on the integration of all necessary ingredients of the QCCD architecture~\textemdash~a scalable trap design, parallel interaction zones, and fast ion transport~\textemdash~into a programmable trapped-ion quantum computer. Using four and six qubit circuits and two parallel interaction zones, the system level performance of the processor is quantified both by the fidelity of a teleported CNOT gate utilizing mid-circuit measurement \cite{Wan19} and a quantum volume~\cite{Cross19} measurement of $2^6=64$. We find that the overall system performance is consistent with the low error rates achieved in the individual ion crystals, demonstrating that the QCCD architecture is a viable path towards high performance quantum computers.

The first quantum logic gates were performed with trapped ions \cite{Cirac95,Monroe95}, and since then researchers have demonstrated coherence times as well as gate, state preparation, and measurement fidelities that are among the best of any viable quantum computing platform \cite{Wang17,Gaebler16,Ballance16,Christensen19}. Recently, researchers have focused their attention towards scaling to larger qubit numbers needed for complex quantum algorithms. While small trapped-ion quantum computers only require a single ion crystal in a single trapping region, this approach is unlikely to scale beyond one hundred qubits \cite{Murali20}. Efforts have therefore focused on architectures that use either a network of ion crystals connected through photonic links \cite{Monroe14,Hucul15} or physically transporting ions between different crystals  \cite{Home09a,Kaufmann17} and eventually even across trap modules \cite{Lekitsche17}.  In this article we demonstrate progress towards the latter approach.

The QCCD architecture aims to create a high fidelity, scalable quantum computer, at the cost of some challenging requirements: \textbf{1} -- a device that can trap multiple small ion crystals capable of high fidelity operations, \textbf{2} -- fast transport operations for moving ions between crystals, \textbf{3} -- tracking of qubit phases and synchronization of the control signals across multiple regions, \textbf{4} -- the likely need for trapping two different ion species, one as a qubit and another to sympathetically cool the crystals back to near the motional ground state after transport, and \textbf{5} -- parallelization of transport and quantum operations across the device. Many of these difficulties have been addressed individually \cite{LabaziewiczThesis08,Maunz16,Bowler12,Kaushal20,Barrett03}, but combining these features into a single machine creates performance requirements that are difficult to reconcile: for example, high fidelity qubit operations require minimal motional excitation and thus low electrode-voltage noise, while fast transport requires high bandwidth voltage controls for those same electrodes. Previous efforts, e.g. refs.  \cite{Home09,Kaufmann17,Wan19}, have demonstrated impressive progress towards a scalable QCCD based quantum computer but have lacked multiple parallel operation zones, sympathetic cooling, or were limited to a single qubit pair.

In this article, we report on the most advanced realization of a QCCD based quantum computer to date encompassing all of the above capabilities. Using up to six qubits, the system operates with low error rates even for deep circuits comprising many transport and gating operations achieving the maximal quantum volume\cite{Cross19} allowed by six qubits without being limited by errors. In addition, we demonstrate small crosstalk errors and that the system is capable of mid-circuit measurements and conditioned feedback~\cite{Nielsen00}. With relatively straightforward upgrades to the optical delivery and ion transport sequences, our architecture will be able to accommodate more qubits.

\begin{figure} 
  \includegraphics[width=0.46\textwidth]{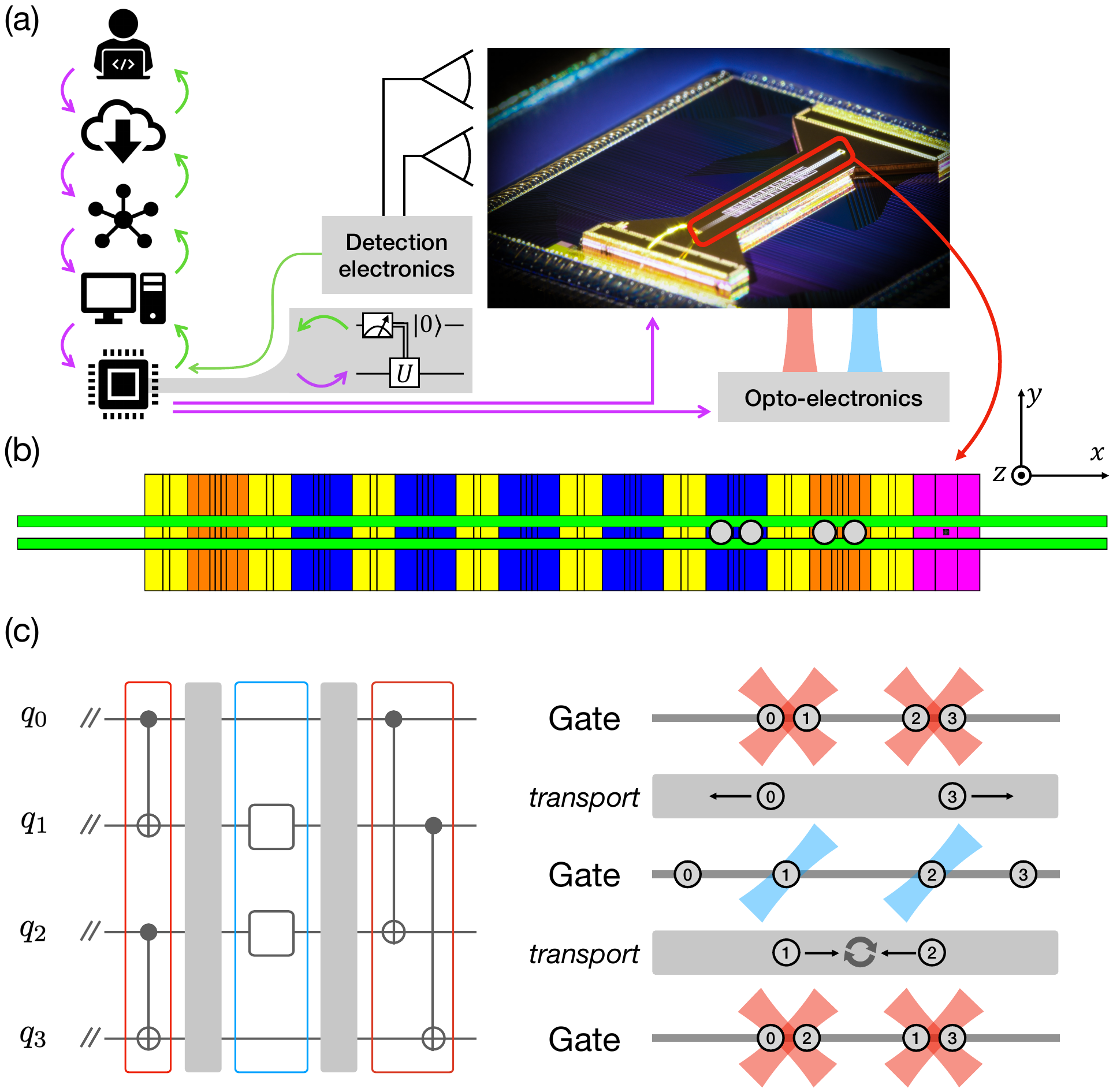}
  \caption{The programmable QCCD quantum computing system. (a) Right, a picture of the trap. Left, the information flow from the user to the trapped ion qubits. From top to bottom, we illustrate: user, cloud, internal tasking, machine control system, FPGA. The circuits are processed by a compiler to generate control signals (purple) sent to both the trap and the optoelectronic devices controlling the laser beams. An imaging system and PMT array collect and count scattered photons, and the results (green) are sent back to the software stack and user, or processed for real-time decision making. (b) A schematic of the trap: RF electrodes (green), loading hole (black), load zone (pink), extended gate zones (orange), gate zones (blue), and auxiliary zones (yellow) for qubit storage. In this work, only the gate zones with grey circles are used.  (c) A general quantum circuit: ions already sharing a gate zone are gated, then spatially isolated for SQ gates, then the second and third ions are swapped for the final TQ gates. While not shown, readout, TQ gates, and SQ gates can all be performed in parallel across different zones.}
\label{fig:arc}
\end{figure}

Requirement \textbf{1} of the architecture is the simultaneous trapping of multiple ion crystals capable of high fidelity quantum operations. The ion trap presented in this work (Fig.~\ref{fig:arc}) has a linear geometry set by RF electrodes (Fig.~\ref{fig:arc}b) that provide a uniform radial trapping force in the $y$ and $z$ directions along a line (called the RF-null) $70 ~\mu$m above the surface (coordinate axes shown in Fig.~\ref{fig:arc}b). Additional trapping potentials and transport capabilities are provided by the 198 independently controlled DC electrodes. The geometry is flexible enough to admit different modes of operation, but we limit our description to the case where the trap is understood as consisting of sixteen different \textit{zones} identified in Fig.~\ref{fig:arc} as gate, extended gate, auxiliary and loading zones. Only the gate and extended gate zones, which are identical in size but with differing numbers of electrodes, are suitable for complex transport operations, making them the most convenient for quantum operations. Here, we use the extended gate and gate zones closest to the loading hole for convenience (Fig. \ref{fig:Beta_Trap}) and refer to them as gate zones 1 and 2 respectively. All quantum operations are executed in these gate zones using lasers propagating parallel to the trap surface. The beams are focused to a spot size small enough to address a single zone but not small enough for individual addressing within a single ion crystal. This means that entangling two-qubit (TQ) gates between arbitrary pairs, and any single-qubit (SQ) gate, measurement, or reset all require ion rearrangement.

In this article we limit the modes of operation to either $N=4$ or $N=6$ qubits, always matching the number of $^{171}$Yb$^{+}$ qubit ions to $^{138}$Ba$^{+}$ coolant ions. The machine is initialized with one (Ba-Yb-Yb-Ba) crystal in each of the two gate zones for the $N=4$ mode and with additional (Yb-Ba) pairs in the load and first auxiliary zones for the $N=6$ mode. 

A quantum computation begins with a user submitting a program in the form of a quantum circuit. A compiler assigns qubits to physical ions so as to minimize the number of transport operations needed and the circuit proceeds as a series of transport and gating operations, pairing and isolating qubits appropriately for SQ and TQ gates, and measurements. After completing all of the operations, the ions are returned to the initial configuration so that the circuit can be repeated to gather measurement statistics before sending the results back to the user.

Transport (requirement \textbf{2}) is achieved through dynamic voltage waveforms applied to the DC electrodes, and falls into three categories:

\textbf{Linear transport:} A potential well is moved along the RF-null from one position to another \cite{Palmero14}.  During a quantum computation, these operations are done in parallel with multiple wells at different positions moving simultaneously. Linear transports always move Yb-Ba pairs, except during trap loading procedures.

\textbf{Split/combine:} A split operation divides a single potential well into two~\cite{Home06}, splitting a 4-ion crystal into two 2-ion crystals. The combine operation is the time reversal of the split operation, and both occur only in gate zones. 

\textbf{Swap:} The qubit order of a 4-ion crystal is flipped by rotating the qubits about an axis perpendicular to the RF-null \cite{Splatt09}. These operations only occur in gate zones.
\

As described in the Methods section (Table \ref{tab:Transport Library}), the individual transport operations have durations ranging from 50 to 300 $\mu$s and introduce a moderate amount of motional excitations to the ion crystal comparable to Doppler temperatures, enabling subsequent sideband cooling routines. Using the three transport primitives, qubits are appropriately arranged for gating using a parallel bubble (odd-even) sorting algorithm, which sorts an array of length $N$ in $\mathcal{O}(N)$ steps \cite{Haberman1979}. The sorting sequences can involve several transport operations for every quantum operation (illustrated in Fig. \ref{fig:N=4_Dance}), but transport failure events are both rare and automatically detected, allowing the data to be scrubbed and uncorrupted (see Methods).
%t

Tracking the phases of the qubits across the spatially separated gate zones (requirement \textbf{3}) requires both stable qubits and phase-stable gate operations. To this end, we use the hyperfine ``clock" states in the $^2S_{1/2}$ ground state of $^{171}$Yb$^{+}$ as the qubit and define $\ket{0}\equiv|F=0, m_F=0\rangle$ and $\ket{1}\equiv|F=1, m_F=0\rangle$, where $F$ and $m_F$ are the quantum numbers for total angular momentum and its z-projection. These states benefit from a relatively low B-field sensitivity and we measure a spin-echo T2 of 2 s in both interaction zones, limited by magnetic field fluctuations.

Our native gate set consists of four types of phase stable gates (see Methods for more details):

\textbf{SQ} $\mathbf{Z}$ \textbf{rotations:} The SQ rotation around $Z$ is done entirely in software by a phase-tracking update as described in the Methods section. We set a resolution limit of $\pi / 500$ for $Z$ rotations.

\textbf{SQ }$\mathbf{X}/\mathbf{Y}$ \textbf{rotations:} Stimulated Raman transitions using co-propagating laser beams on an isolated Yb-Ba crystal apply SQ rotations about an arbitrary axis in the $X$-$Y$ plane of the Bloch sphere. We currently restrict our gate set to $\pi$ and $\pi/2$ rotations. Arbitrary SQ gates submitted by the user are synthesized into at most two $X/Y$ rotations along with an additional $Z$-rotation. Because the Raman beams are co-propagating, the angle of the rotation axis is set by the phase of the microwave beatnote of the laser beams and not affected by optical path length drifts, allowing for straightforward synchronization across zones.

\textbf{TQ gates:} We operate a M{\o}lmer-S{\o}rensen interaction \cite{Molmer00} in the phase-sensitive configuration \cite{Lee05,Baldwin19} and add SQ wrapper pulses driven with the same lasers (Fig.~\ref{ZZ_gate}) to remove the optical phase dependence, generating the phase-insensitive entangling gate $U_{zz}=\text{exp}(-i\frac{\pi}{4}Z\otimes Z)$. Mapping the gate to the $Z$ basis allows the operations to be performed in parallel across multiple zones without the need for additional phase stabilization or synchronization.

\textbf{Global microwave rotations:} A microwave antenna in the vacuum chamber applies global qubit rotations with a variable phase. 
The microwave field amplitude has a $3\%$ inhomogeneity across the different gate zones and, therefore, is not suitable for logical operations, nor is it available to be used in quantum circuits. However, microwaves are used to suppress memory errors through dynamical decoupling \cite{Viola99}. 
Since the amplitude is inhomogeneous, we cannot apply global microwave $\pi$ pulses, as is typical for dynamical decoupling. 
To avoid the accumulation of coherent errors, we apply pulses in pairs with opposite phases during ground-state cooling without any transport operations in between, canceling out the small amplitude errors while preserving most of the memory error suppression benefits.

Qubit initialization and measurement are done by spatially isolating a single Yb-Ba crystal so the resonant light has a minimal effect on the other qubits, allowing measurement (and subsequent reinitialization, if necessary) to be performed at any point in the circuit. Standard optical pumping and state-dependent fluorescence procedures are used for initialization and measurement~\cite{Olmschenk07} (see Methods).

%\
Aside from the global microwave operations, quantum operations are characterized in both gate zones using RB (randomized benchmarking) techniques~\cite{Magesan11}. The similar environments for the two zones result in nearly identical operation times and fidelities, summarized in Table \ref{tab:benchmarks} and Table \ref{tab:Qubit_Operation_Times}. An estimated error budget for TQ gates is given in Table \ref{tab:error_budget} in the Methods section.

%\
\begin{table*}
\footnotesize
\begin{tabular}{|l|l|l|l|l|}
\hline
Component          & Zone Avg. & Zone 1 & Zone 2 \\ \hline
SPAM (simultaneous SQ RB)                &    $3(1)\times10^{-3}$       &    [$3(1)\times 10^{-3}$, $3(1)\times 10^{-3}$]    &  [$2(1)\times 10^{-3}$, $3(1)\times 10^{-3}$]      \\ \hline
SQ gates (simultaneous RB)   &  $1.1(3)\times10^{-4}$  & [$1.4(3)\times 10^{-4}$, $1.2(2)\times 10^{-4}$]  &  [$9(2)\times 10^{-5}$, $1.0(3)\times 10^{-4}$]   \\ \hline
TQ gates (individual RB)       &   $7.9(4)\times10^{-3}$    & $6.7(5)\times10^{-3}$    &    $9.0(4)\times10^{-3}$     \\ \hline
TQ gates (simultaneous RB)     &   $8.0(4)\times10^{-3}$        &    $7.2(4)\times10^{-3}$    &   $8.8(4)\times10^{-3}$     \\ \hline
\end{tabular}
\caption{Component benchmarking results. Numbers are the average error per operation. For SPAM and SQ gates, the two numbers represent the measured values for each qubit in the corresponding zone using the high-fidelity (120 $\mu$s duration) detection.}
\label{tab:benchmarks}
\end{table*}
%\

We meet the sympathetic cooling requirement (\textbf{4}) with $^{138}$Ba$^+$ as the coolant ion and employ both Doppler and resolved sideband cooling \cite{MonroeCooling95} at various times throughout a quantum circuit, (see Methods section for details). The cooling laser light is tuned near the $^{138}$Ba$^+$ transition at 493.5 nm, far enough from any resonances in $^{171}$Yb$^+$ to prevent induced errors. These cooling protocols are the current runtime bottlenecks in our device (Fig. \ref{fig:N=4_Dance}b , Fig. \ref{fig:CNOT_time_budget}, Fig. \ref{fig:QV_time_budget}) and likely necessary for the QCCD architecture but can be significantly improved~\cite{Jordan2019}.

The final requirement (\textbf{5}) of parallelized quantum operations and transport is verified at the system level through holistic benchmarks. We performed three holistic measurements aimed at characterizing the system performance as a general quantum information processor: simultaneous RB in the different gate zones, a teleported CNOT gate using mid-circuit measurements, and a measurement of the system's quantum volume.
%\

We characterize the performance of the SQ and TQ gates (absent of interzone transport) via RB. We also generalize the SQ simultaneous RB method~\cite{Gambetta12} to TQ operations to detect crosstalk errors. As described in the Methods section, this benchmark compares the results of RB tests performed separately in each zone versus simultaneously in the two zones. As summarized in Table \ref{tab:crosstalk} and Fig. \ref{fig:correlations}, we find the crosstalk errors to be $<0.1\times10^{-3}$ per gate and consistent with zero to within the measurement uncertainty, highlighting the benefit of one the central design features of the QCCD architecture: using spatially separated gate zones addressed by localized laser beams.

Mid-circuit measurements that preserve the quantum information stored in other qubits have been achieved in ion trap quantum computing experiments by splitting and spatially isolating ion crystals ~\cite{Barrett04} as well as by using an ancilla ion species ~\cite{Negnevitsky18, Wan19}. Our architecture uses the former approach and we demonstrate this capability at the system level by performing a teleported CNOT gate circuit as in Ref.~\cite{Wan19}.  Quantum gate teleportation is a protocol for applying a gate between a pair of remote qubits~\cite{Plenio2000} requiring entangling operations, mid-circuit measurements, and classically-conditioned quantum gates - all of which are necessary for quantum error correction.  The protocol uses four qubits and through a combination of three entangling operations, two mid-circuit measurements and two measurement-conditioned SQ operations, executes a quantum CNOT gate between two qubits that never directly interact with each other (see Methods). By evaluating all four input states in two different bases (eight total input states), we bound the fidelity of the operation to be $F_{\rm avg}\ge0.899(6)$.

\begin{figure}
\begin{centering}
  \includegraphics[width=0.96 \columnwidth]{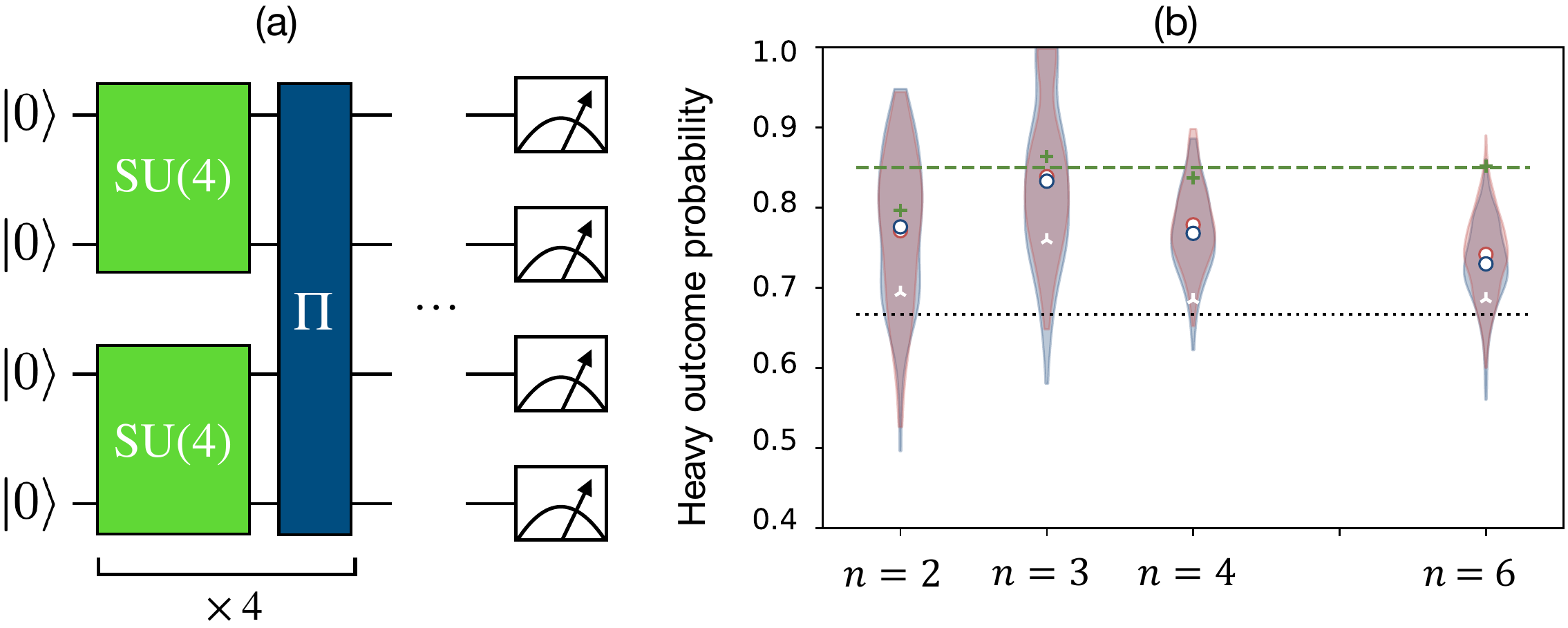}
  \caption{QV circuits and measurements. (a) The test interleaves random SU(4) gates, each requiring three entangling gates, with random permutations ($\Pi$) of qubits (this example illustrates the $N=4$ case). (b) Results for QV test for $N=2,3,4,$ and $6$. Each test was performed with randomly generated circuits from Qiskit~\cite{Qiskit} with the blue violin plots showing the proportion of circuits with a certain heavy outcome probability given by the horizontal width of the plot. Noisy simulations run in Qiskit with depolarizing errors rates estimated from RB experiments are plotted as violin plots in red. Green crosses give the average of the ideal outcomes for the circuits run. The green dashed line is the average over all ideal circuits and the black dashed line shows the passing threshold. The white tri-points show the two-sigma points in the measured distributions.}
\label{fig:qv}
\end{centering}
\end{figure}

Quantum volume (QV) quantifies the effective power of a quantum computer by measuring its ability to produce quantum states that are hard to classically simulate for large qubit numbers. The test is not meant to mimic any particular quantum algorithm; rather, it is meant to stress several metrics thought to be important for general purpose quantum computers (including qubit number, gate fidelity, crosstalk, connectivity, and flexibility in native gate sets) and provides a single number to gauge the general potential of the system. A QV test for $N$ qubits consists of both experimental runs and classical simulations of $O(N)$ depth circuits (illustrated in Fig.~\ref{fig:qv}a) that require a quadratic number of SQ and TQ gates ($3 N^2/2$ total TQ gates between randomly paired qubits) and resemble important variational algorithms~\cite{McClean16,Farhi14}. The experimentally measured outputs, together with the classically simulated ideal outputs, are evaluated against the heavy-outcome criteria (a criteria ensuring that the measured outcomes reproduce the ideal distribution sufficiently better than random noise~\cite{Aaronson16, Cross19}). If the circuits pass this test $2/3$ of the time with two-sigma confidence, the system is said to have $\text{QV}=2^N$, a number which approximates the complexity of the classical simulations in the test. The classical simulations required for QV limit its usefulness to $N< 50$ qubits, but the test is challenging even for small numbers of qubits and establishes worthwhile goals for gates and system level performance~\cite{Cross19} for near-term devices.

\begin{figure}
\begin{centering}
  \includegraphics[width=0.75 \columnwidth]{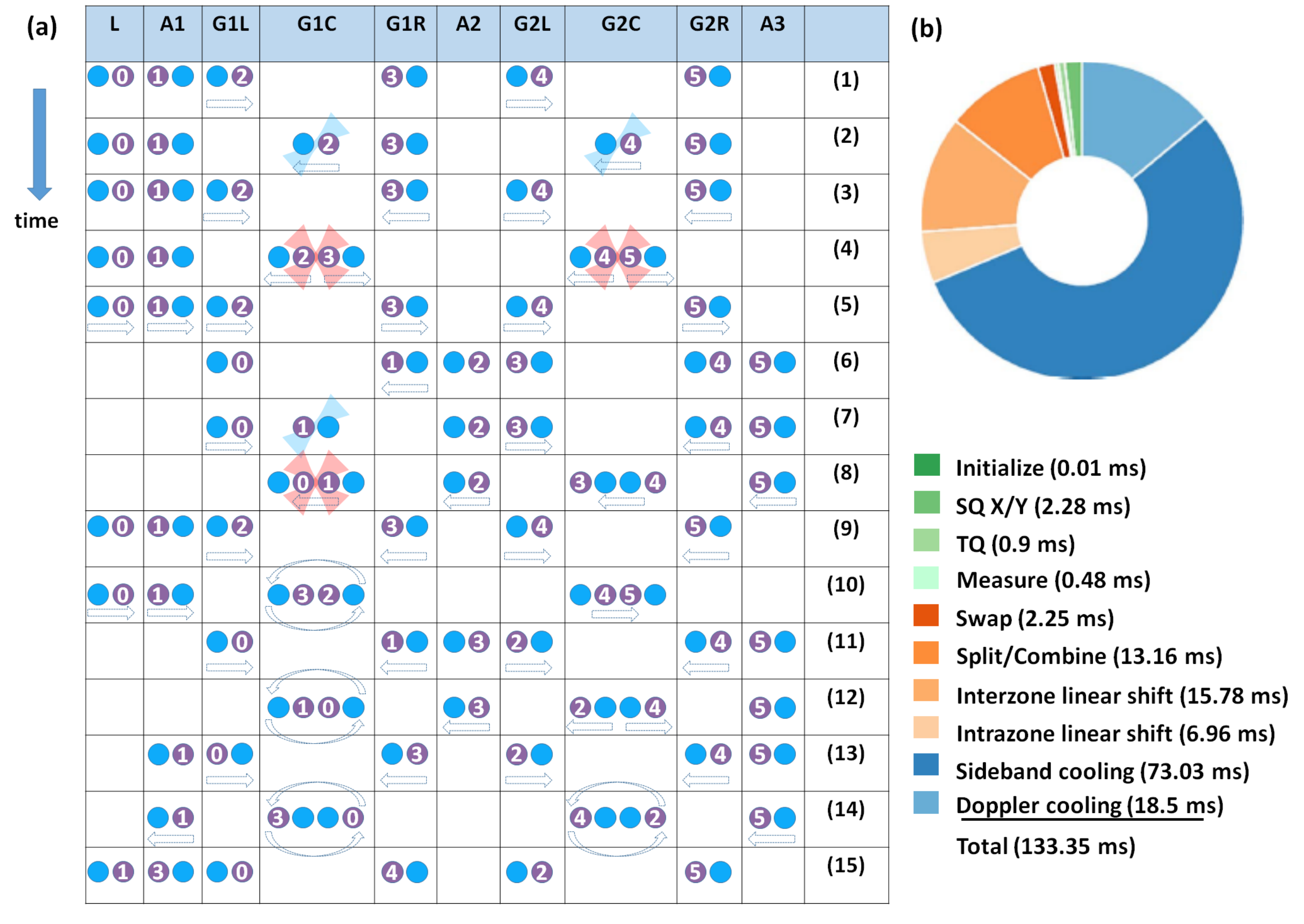}
  \caption{Transport operations and timing for an $N=6$ QV test circuit. (a) An example of the transport operations between two SU$(4)$ gates with the random qubit permutation $\Pi=(0,1,2,3,4,5)\rightarrow(1,3,0,4,2,5)$ (coolant ions are not numbered and cooling steps are not called out). Similar transport sequences are repeated approximately 6 times (except rare cases that gate the same qubits twice in a row) to achieve circuit depth 6.  Trap regions (in gate zones, we can create multiple wells so there can be multiple regions per zone) are labeled as load zone (L), auxiliary zone 1 (A1), gate zone 1 - left (G1L), gate zone 1 - center (G1C), gate zone 1 - right (G1R), auxiliary zone 2 (A2), gate zone 2 - left (G2L), gate zone 2 - center (G2C) and gate zone 2 - right (G2R). The distance from the center of the gate zone to the left and right regions is 110$~\mu$m. The different steps described in the text are labeled as follows: (1) linear shift, (2) SQ gate, linear shift, (3) combine, (4) TQ gate, split, (5) linear shift, (6) linear shift, (7) SQ gate then combine, (8) TQ gate, split, linear shift, (9) combine, (10) swap, split, linear shift, (11) combine, (12) swap, split, linear shift, (13) combine, (14) swap, then linear shift. Operation times are in Tables \ref{tab:Transport Library} and \ref{tab:Qubit_Operation_Times}. (b) Detailing the total time spent on various operations during an $N=6$ QV circuit. The $N=6$ QV circuits have approximately 15 random SU$(4)$ gates, compiling down to 45 TQ gates.}
\label{fig:N=4_Dance}
\end{centering}
\end{figure}

As detailed in the Methods section, we performed QV tests on $N=2,3,4$ and $6$ qubits and passed in all cases: for $N=2$ with 77.58\% of the circuits with heavy outcomes and 99.56\% confidence, $N=3$ with 83.28\% of the circuits with 99.9996\% confidence, $N=4$ with 76.77\% of the circuits and 99.16\% confidence, and $N=6$ with 72.96\% and 99.77\% confidence (results are plotted in Fig.~\ref{fig:qv}b). Note that without errors and in the limit of large $N$ the circuits return $\sim85\%$ heavy outcomes with slight variations for smaller $N$ (Fig.~\ref{fig:qv}) and a completely depolarized circuit (large error limit) has a heavy outcome $50\%$ of the time. Prior to this manuscript, the highest published QV was 16~\cite{Cross19}, and after this manuscript was submitted, QV 64 was also demonstrated by IBM~\cite{Jurcevic20}. Fig. \ref{fig:qv}b also shows theoretical simulations of the circuit outcomes assuming a depolarizing noise channel for the TQ gates, with a noise magnitude extracted from TQ RB.  The excellent agreement demonstrates that the QCCD architecture can fulfill its primary goal: maintaining the gate fidelities achievable in small ion crystals while executing arbitrary circuits on multiple qubits.

We are currently scaling the optical delivery to five gates zones along with improved transport capabilities allowing for increased qubit numbers and circuit complexities. The current runtime bottlenecks for the device are the cooling procedures. This bottleneck can be improved with the combination of improved transport operations \cite{Mourik2020} and cooling techniques \cite{Jordan2019}. The current algorithmic bottleneck is the TQ gate fidelity, which is consistent with error models pointing to voltage noise in the waveform generators and spontaneous emission as the two dominant error sources. These errors can be mitigated with advanced filtering techniques and larger Raman laser detunings, respectively.

The path to truly large systems of thousands or millions of qubits remains unclear. Our architecture could be scaled in a straightforward manner by placing multiple linear chip-traps end-to-end without major changes in the individual components used. This linear scaling would result in a linearly increasing transport time for fully connected circuits, limiting the system's achievable connectivity due to a finite coherence time in the absence of fault-tolerant methods. Other scaling routes that would better maintain connectivity include using flying qubits (photonic interconnects or specially designed high-speed transport structures), or a 2D trap geometry.  The latter approach requires junction transport and a laser geometry that would avoid beam clipping on a large chip~\cite{Mount_2013,Mehta2020}. The question of how to most efficiently move quantum information through a computer is not settled, and the likely need for quantum error correction introduces additional complications. For example, leveraging high connectivity is expected to enable more efficient encodings \cite{Kovalev2013} and the true optimal design may ultimately require different connectivities at different scales.

\appendix
\section{Methods}

\begin{figure} 
\begin{centering}
  \includegraphics[width=1.0\columnwidth]{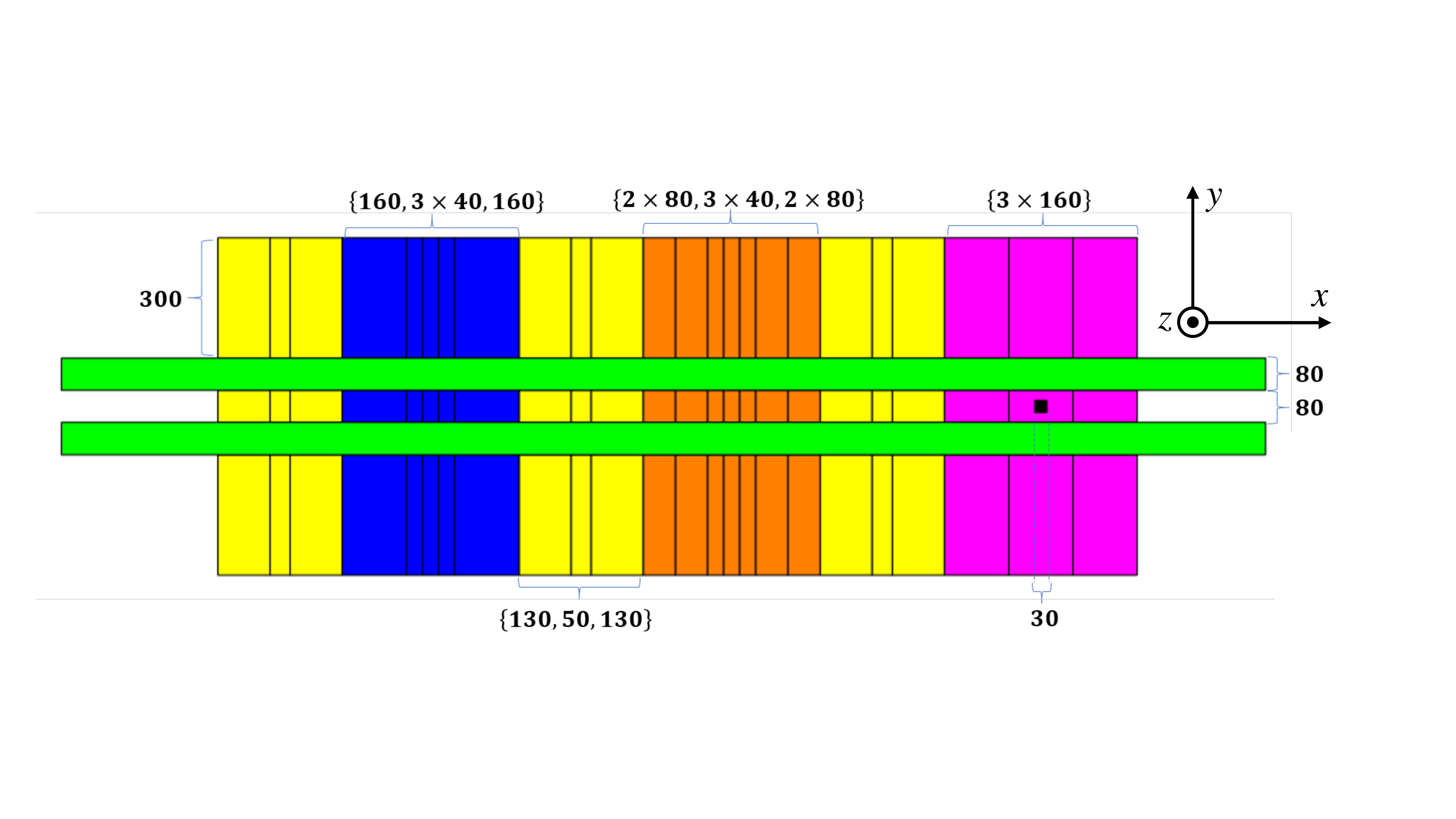} 
  \caption{Close-up schematic of the trap region used in this work. Dimensions are in $\mu$m. We refer to the orange gating region on the right as zone 1 and the blue gating region as zone 2.}
\label{fig:Beta_Trap}
\end{centering}
\end{figure}

\textbf{Honeywell Surface Trap:} We designed and fabricated a 2D surface trap at Honeywell's microfabrication facility in Plymouth, MN, a section of which is shown in Fig. \ref{fig:Beta_Trap}. The 198 DC gold electrodes were fabricated with an undercut etch to mitigate stray fields by eliminating a line of sight between dielectrics and the ions \cite{Maunz16}. The trap is cooled to 12.6 K via a cold finger attached to a liquid He flow cryostat with stability better than 2 mK. Single $^{171}$Yb$^{+}$ axial heating rates at a 0.97 MHz trap frequency vary from 100-500 quanta/sec and typical radial heating rates are $\sim500$ quanta/sec, both limited by technical noise on electrode controls.

\textbf{Computer Control:} The processor is programmed using the quantum circuit model~\cite{Nielsen00}, compiling gate sequences into hardware commands for the machine control system, which is also responsible for circuit queueing and calibration routines. Automated calibrations take up about 25$\%$ of the system duty cycle and are either executed on a schedule or triggered when a measured parameter exceeds the specified tolerance. Clock synchronization between qubits is maintained via a phase-tracking protocol that updates the qubit phases after transport and gate operations to account for AC-Stark shifts and interzone differences in the SQ phase.

Data corruption from ion loss or crystal reordering events (caused by transport failures and background collisions) is mitigated by probing the ion crystal mode structure at the end of a series of circuit runs. When these events are detected, the data since the previous check is discarded and the circuits are repeated. The detection procedure is blind to events that swap the qubit ions as well as multiple reordering events that leave the ions in the correct order immediately prior to the probe.  For the $N=6$ QV test where 100 repetitions of 400 different circuits were run over a period of five hours ($\sim10^7$ transport operations) only three such events were detected. 

\textbf{Trapping and Transport Operations:} A hole in the trap (Fig. \ref{fig:Beta_Trap}) allows neutral atoms from an effusive thermal source to enter the trapping region where they are photoionized and Doppler cooled using standard techniques \cite{Olmschenk07} and then transported to a gate zone.

A 190V RF drive at 42.35 MHz provides radial confinement in the trap and DC voltages provide axial confinement ($x$-direction). The radial principal axes are at 45 degrees relative to the trap surface. The single Yb ion trap frequencies are $\{\omega_{\tilde{x}},\omega_{\tilde{y}},\omega_{\tilde{z}}\}=2\pi\times\{0.97,2.8,2.7\}$ MHz, resulting in an 8 $\mu$m long four-ion crystal. The principle axes are related to the trap axis coordinates in Fig.~\ref{fig:Beta_Trap} as $\{x,y,z\}=\{\tilde{x},\frac{\tilde{y}+\tilde{z}}{\sqrt{2}},\frac{\tilde{y}-\tilde{z}}{\sqrt{2}}\}$.

The transport primitives in Table \ref{tab:Transport Library} are constructed using methods similar to those in Ref.~\cite{Blakestad10}. Waveforms are solved for with a quadratic program minimizing the voltage amplitudes while generating a potential whose derivatives are constrained in space and time. Specifically, linear transport programs specify a trap minimum moving along the RF-null, swaps specify the orientation of the principal axes and confinement strengths through time, and the split-combine operation is solved for by specifying a quartic potential term and its strength relative to a quadratic term.  During the split operation, the quadratic terms changes from positive to negative, creating a double-well potential.  The linear shifts and split-combines work for both individual species, as well as mixed-species crystals, with no modifications.  The swaps were calculated for a specific crystal orientation (BYYB or YBBY).

Transport sequences are determined using a heuristic algorithm to assigning gating operations to specific zones and times. Individual transport primitives in the library specify dynamic voltages applied to the electrodes via arbitrary waveform generators. The waveforms are stitched together using an interpolation procedure to ensure voltage continuity. The superposition principle for voltages along with the linearity of the constraints guarantees that a linear interpolation between trapping potentials also yields a valid trapping potential.

\begin{table}
%\footnotesize
\begin{tabular}{|p{2.0cm}|p{2.0cm}|p{2.2cm}|p{2.2cm}|}
\hline
 Operation  &Duration ($\mu$s)  &Axial heating (quanta/mode) &Radial heating (quanta/mode)  \\ \hline

Intrazone shift  &58  &$<$1  &$<$0.5  \\ \hline
Interzone shift  &283  &$<$1  &$<$0.5 \\ \hline
 Split/combine  &128  & $<$1  &$<$0.5 \\ \hline
 Swap  &200  &$<$2  &$<$2 \\ \hline
\end{tabular}
\caption{Transport primitive library and associated durations and heating. For heating estimates we fit four ion crystal spin-flip data to a model that assumes all modes are at the same temperature. The fitted temperature increases are converted to units of quanta/mode (note that the inequality holds for all modes). The times shown do not include interpolation between different operations or small delays in the electronics, which increase the time for every operation by $\sim10\%$. The interzone shift is a linear shift between the two gate zones, while an intrazone shift moves ions within a single gate zone by 110 $\mu$m for single qubit addressing.}
\label{tab:Transport Library}
\end{table}

Transport waveforms are constructed using the electric field for each electrode computed with a 3D electrostatic simulation of the trap, and shimmed to account for stray electric fields. Shim voltages are determined using calibration routines and added to the trapping voltages in gate zones during the gate, split/combine and swap operations.  Successful split/combine requires the axial shim to be set to a few V/m precision, typically $<$ 30 V/m, and fluctuates by less than 5 V/m throughout a day.  The $y$ and $z$ shims are typically $<$ 50 V/m and $<$ 500 V/m respectively. Over the course of a day, the radial shim field can vary by a few tens of V/m, but the system's performance degrades when off by $\gg$ 10 V/m.

\textbf{Qubit Manipulation:} A 5G magnetic field is applied in the xy plane at $45^{\circ}$ with respect to the x-axis, uniform between the zones to within 0.2 mG, matching qubit frequencies to within 1 Hz. The qubit coherence time is characterized by a single-pulse spin-echo sequence in the two zones, showing 1/e Gaussian decay times of $2.0(2)$ s and $2.0(3)$ s for the first and second gate zones respectively. Lasers are directed into the gate zones at either 45 or 90 degrees to the trap axis with a beam diameter of $\sim17 ~\mu$m and the laser power budget is dominated by the TQ gate beams with 8.5 mW per beam.

\begin{table}
%\footnotesize
\begin{tabular}{|l|c|}
\hline
 Operation  &Duration($\mu$s)  \\ \hline
Qubit initialization  &10    \\ \hline
Qubit measurement (high-fidelity)  &120   \\ \hline
Qubit measurement (low crosstalk)  &60   \\ \hline
Cooling stage 1 (Doppler)&550   \\ \hline
Cooling stage 2 (Axial and Radial SB)&850   \\ \hline
Cooling stage 3 (Axial SB)&650   \\ \hline
SQ $\pi/2$ time  &5  \\ \hline
TQ gate & 25\\ \hline
\end{tabular}
\caption{Qubit manipulation times. Circuits can be run using two different measurement protocols. For circuits where all measurements are made at the end, we use the high-fidelity measurement setting. Circuits containing mid-circuit measurements use shorter duration measurements to minimize the crosstalk error on idle qubits. The shorter detection time measurement error is $\sim7\times10^{-3}$, about twice as large as those reported in Table \ref{tab:benchmarks}. Mid-circuit measurements induce an error of $\sim1\%$ on neighboring idle qubits as measured by a Ramsey experiment. There are three different cooling stages used during transport (stages 1 and 2) and before gates (stage 3) and are either implemented through Doppler or SB (sideband) cooling.}
\label{tab:Qubit_Operation_Times}
\end{table}

%\
\textbf{State preparation and measurement (SPAM):} The qubits are initialized and measured using standard optical pumping and state-dependent fluorescence techniques \cite{Olmschenk07}, and the ions' fluorescence is imaged onto a standard multichannel PMT array for measurements. Detection is performed at the gate zone centers and we set the magnification of the imaging system such that the zone centers line up with every other PMT channel. 

The SPAM errors with 120 $\mu$s detection time listed in Table \ref{tab:benchmarks} are extracted from the $y$-intercept of the survival probability in SQ randomized benchmarking (RB).  Theoretical modeling suggests the state preparation error is $\leq10^{-4}$, and the photon collection efficiency of $\sim3\%$ limits the measurement error to $>10^{-3}$. During these operations, idle qubits are held at least 110 $\mu$m away to avoid decoherence from scattered light. Measurement crosstalk errors are measured to be 0.35$\%$-1.5$\%$ as explained in the caption of Table \ref{tab:Qubit_Operation_Times}, while calculated errors from the detected ion's fluorescence at this separation are $<10^{-3}$. A perfect Gaussian beam 110 $\mu$m away gives a $<10^{-4}$ error suggesting the current limitations are due to imperfect beam profiles and scattering off the trap surface.

\textbf{Gating operations:} Quantum logic operations use stimulated Raman transitions in two different configurations: SQ gates use pairs of co-propagating 370.3 nm Raman beams with circular polarization, and TQ gates use pairs of linearly polarized beams that couple to an axial mode of motion (along the RF-null).

\begin{figure*} [!t]
  \includegraphics[width=1.0\textwidth]{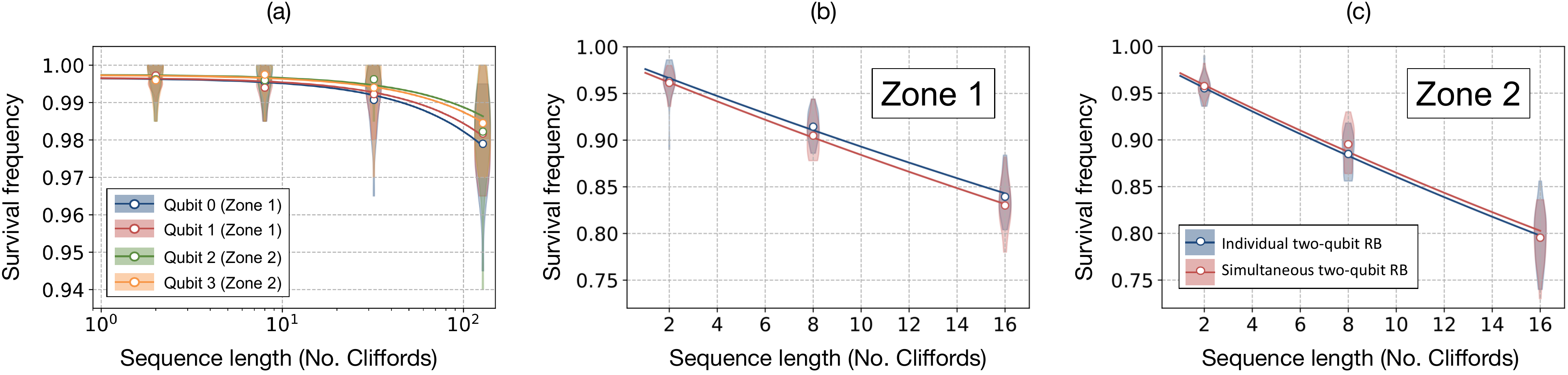}
  \caption{RB for $N=4$ qubits. (a) SQ RB for all four qubits. (b,c) TQ RB results for both zones. We performed TQ RB in each zone, both separately and simultaneously, and observe no statistically significant difference between the two data sets, indicating that crosstalk errors do not play a significant role in our TQ gates.}
\label{fig:all_rb}
\end{figure*}

SQ gates are characterized by performing SQ RB with each qubit following the standard Clifford-twirl version of RB outlined in Ref.~\cite{Magesan11}, with results plotted for each qubit in Fig.~\ref{fig:all_rb}a. Likewise, TQ gates are characterized with TQ RB in each zone independently and then in both zones simultaneously~\cite{Magesan11}. Results are plotted in Fig.~\ref{fig:all_rb}(b,c), and we report the average infidelity of the TQ Clifford gates scaled by the average number of $U_{zz}$ gates per Clifford gate (1.5). The laser pulses that drive the M{\o}lmer-S{\o}rensen interaction use a $\textrm{sin}^2(\frac{\pi}{2}t/T)$ envelope with $T=1.5~\mu$s for turning on and off to suppress unwanted spectator mode excitations. As indicated in Table \ref{tab:error_budget}, our models predict that TQ gate errors are dominated by position fluctuations of the ions stemming from voltage noise, introducing an effective phase noise source. Our estimate for the magnitude of this error uses the measured voltage noise spectral density and a numerically calculated phase noise filter function \cite{Biercuk_2011}.

\begin{figure} 
\begin{centering}
  \includegraphics[width=0.8\columnwidth]{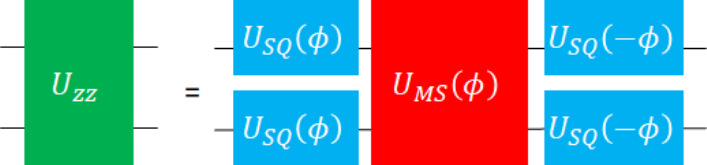}
  \caption{Construction of a phase-insensitive TQ gate. The M{\o}lmer-S{\o}rensen interaction generates the unitary $U_{MS}=\text{exp}(-i\frac{\pi}{4}(X\text{sin}\phi+Y\text{cos}\phi)^{\otimes2})$ (red) whose basis is determined by the optical phase $\phi$. SQ operations driven by the same laser beams generate the unitary $U_{SQ}=\text{exp}(-i\frac{\pi}{4}(X\text{cos}\phi+Y\text{sin}\phi))$ (blue) and are applied globally to both qubits. The resulting composite gate is, up to a global phase, given by $U_{zz}=\text{exp}(-i\frac{\pi}{4}Z\otimes Z)$ (green).}
\label{ZZ_gate}
\end{centering}
\end{figure}

\begin{table}
\footnotesize
\begin{tabular}{|l|c|}
\hline
Error Source          & Magnitude($10^{-3}$)\\ \hline
Spontaneous emission                &    $1.5$  \\ \hline
Debye-Waller   &  $0.1$   \\ \hline
Trap frequency fluctuations       &   $<0.1$   \\ \hline
Laser phase noise     &   $<0.5$     \\ \hline
Heating     &   $<1.0$     \\ \hline
Spectator mode coupling & $0.1$ \\ \hline
Position fluctuations (voltage noise induced)    &   $5$   \\ \hline
Total	& $<8.2$ \\ \hline
\end{tabular}
\caption{An estimated error budget for a single TQ gate operation.}
\label{tab:error_budget}
\end{table}

\textbf{Simultaneous RB}
This method has two parts: (1) identify differences in RB decay rates when applying gates individually versus simultaneously, and (2) identify correlated errors when applying gates simultaneously (also similar to Ref.~\cite{Harper19}). 

TQ simultaneous RB is performed by three experiments: (1) RB in zone one, (2) RB in zone two, and (3) RB in both zones simultaneously. For reference, we fit the RB data to the decay equation $p(\ell) = A \alpha^{\ell} + B$ where $p(\ell)$ is the average survival frequency for a length $\ell$ RB sequence, $A$ is the SPAM parameter, $\alpha$ is the RB decay rate, and $B$ is the asymptote, which is fixed by randomizing the final measurement and the uncertainty is calculated via a semi-parametric bootstrap resample~\cite{Meier06,Baldwin19}.

Addressability errors $\gamma_z$ induced in one zone from operations in the other zone are quantified as the difference between the decay rates from experiments one and two, $\alpha_z$ for $z$ = 1, 2, and the decay rates from experiment three, $\alpha_z^{\textrm{both}}$~\cite{Gambetta12},
\begin{equation}
\gamma_{z} = | \alpha_z - \alpha_z^{\textrm{both}}|.
\end{equation}

As summarized in Table~\ref{tab:crosstalk}, the measured values of $\gamma_z$ are below $10^{-3}$ and within one standard deviation of zero.

\begin{table}
\footnotesize
\begin{center}
\begin{tabular}{|l|l|l|l|l|}
\hline
Zone & $\alpha_z$ & $\alpha_z^{\textrm{both}}$ & $\gamma_z$\\ \hline
1 & 0.9866(9) & 0.9856(8) & $9(10) \times 10^{-4}$ \\ \hline
2 & 0.9820(8) & 0.9824(8) & $4(12) \times 10^{-4}$ \\ \hline
\end{tabular}
\caption{\label{tab:crosstalk} Addressability error estimates for simultaneous RB. The quantities $\alpha_z^{\textrm{both}}$ and $\gamma_z$ refer to simultaneous RB decay rates and their deviation from individual RB decay rates as described in the text.}
\end{center}
\end{table}

\begin{figure}[!t]
  \includegraphics[width=0.5\textwidth]{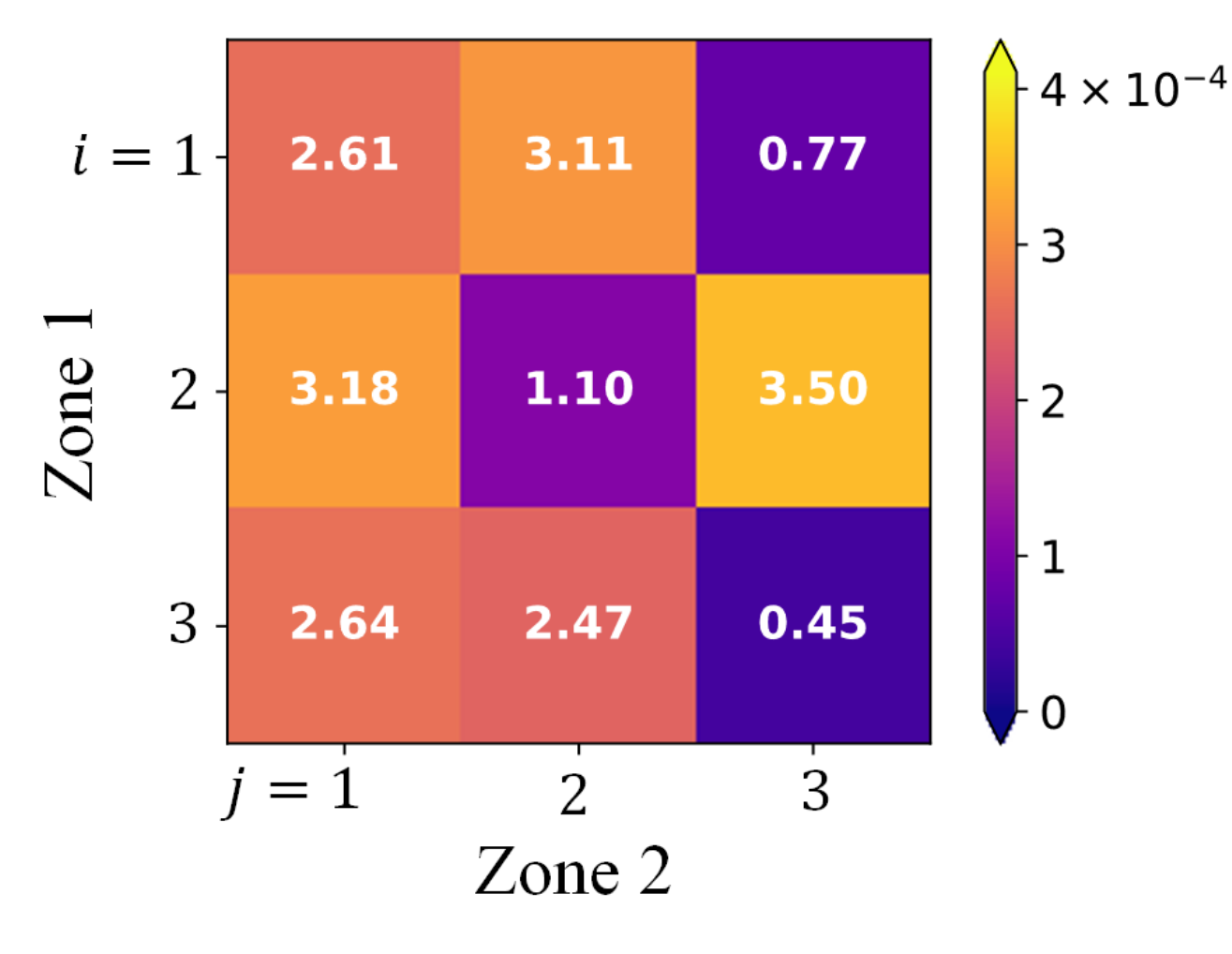}
  \caption{Correlation parameters for simultaneous TQ RB. Each square represents a value of $\delta_{i,j}$ with the numbers being scaled by $10^{-4}$. All are within one standard deviation of zero.}
\label{fig:correlations}
\end{figure}

Correlation errors shared between zones are quantified by running RB simultaneously, first developed for SQ gates~\cite{Gambetta12}. For SQ gates, the structure of the two-subsystem Clifford group $\mathsf{C}_1^{\otimes 2}$ leads to three decay rates, two from individual errors on each qubit and the other from correlated errors, estimated by measuring the Pauli observables $\langle \mathds{1} Z \rangle$, $\langle Z\mathds{1} \rangle$, and $\langle Z Z \rangle$. For parallel TQ gates, there are still three decay rates for $\mathsf{C}_2^{\otimes 2}$ but there are fifteen different Pauli observables. Six of these observables decay due to errors occurring separately on the two pairs of qubits. These are the Pauli operators with the identity acting on one of the qubit pairs, and at least one Pauli Z acting on the other qubit pair, labeled as $\beta_{z, i}$ for $z=1,2$ and $i=1,2,3$. If standard RB assumptions hold, then $\beta_{z,i} = \alpha_{z}^{\textrm{both}}$ for all $i$. The remaining nine observables have at least one Pauli Z acting on each qubit pair and decay due to correlated errors between the two pairs of qubits, labeled as $\mu_{i,j}$. If there are no correlated errors then $\mu_{i,j} = \beta_{1,i} \beta_{2,j}$. The magnitude of correlation errors is quantified by
\begin{equation}
\delta_{i,j} =|\beta_{1,i} \beta_{2,j} - \mu_{i,j}|,
\end{equation}
for all Pauli observables. Fig.~\ref{fig:correlations} shows $\delta_{i,j}$ for all nine values. The minimum uncertainty is $4.1\times10^{-4}$ putting all measured values within one standard deviation of zero.

Simultaneous TQ RB strongly indicates that addressability and correlation errors are minimal in this system. However, our method does not capture all possible crosstalk errors as there may be errors that do not change the decay of an RB experiment but still affect a quantum circuit. We also note that this measurement indicates that crosstalk between gate zones is not a measurable contributor to the gate errors, but other modes of operation in which qubits are stored in auxiliary zones during gate operations require further characterization.

\begin{figure}[!t]
\begin{center}
\includegraphics[width=1.0\columnwidth]{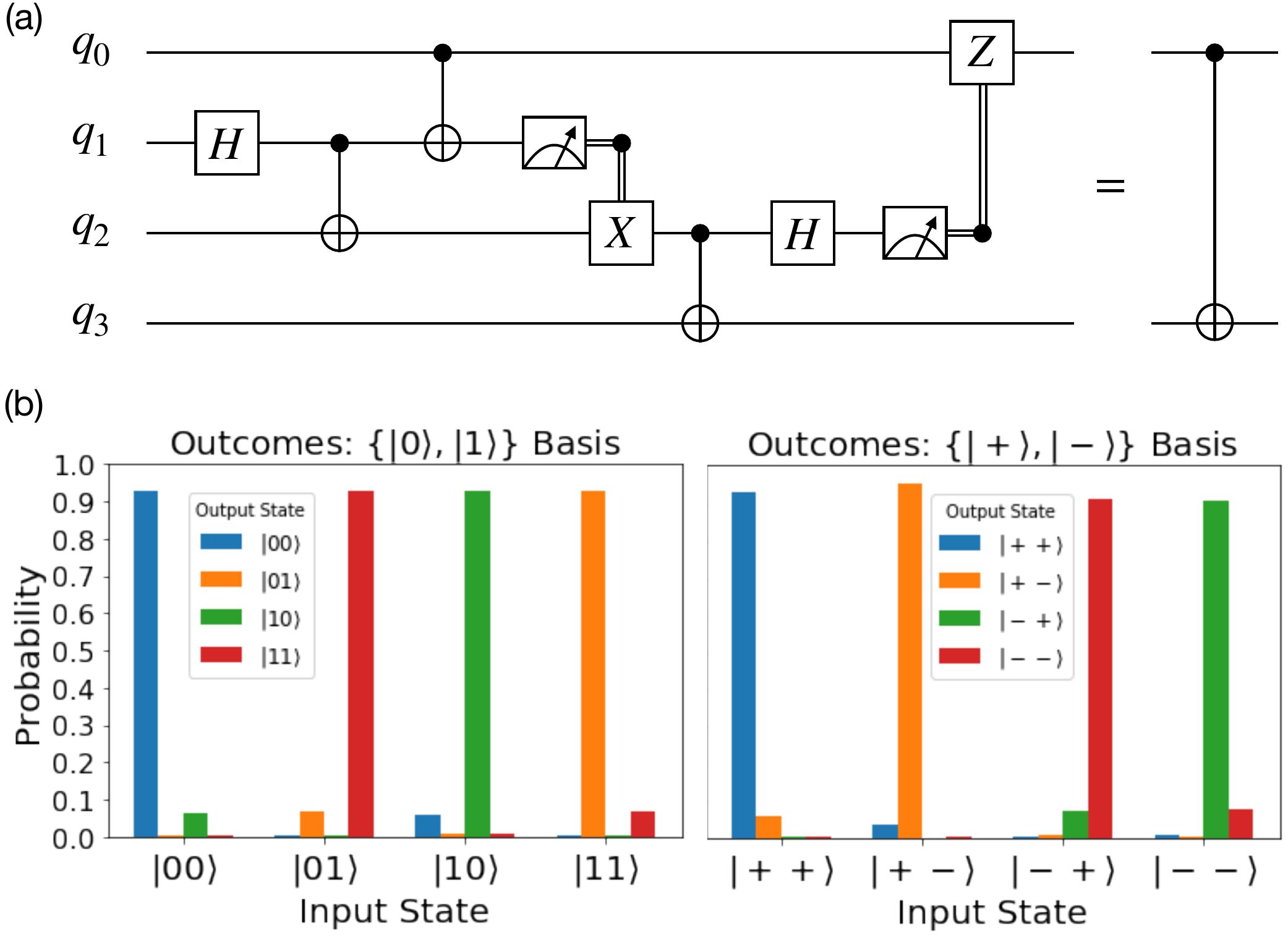}
\caption{Teleported CNOT circuit. (a) Circuit implementing a teleported CNOT gate,
with $q_0$ and $q_3$ the control and target qubits, respectively.
(b) Bar plots showing the distribution of measurement outcomes when qubits
$q_0$ and $q_3$ are prepared and measured in the $\{\ket{0}, \ket{1}\}$ and $\{\ket{+}, \ket{-}\}$ bases.}
\label{fig:CNOT_teleportation}
\end{center}
\end{figure}

\begin{figure}[!t]
\begin{center}
\includegraphics[width=1.0\columnwidth]{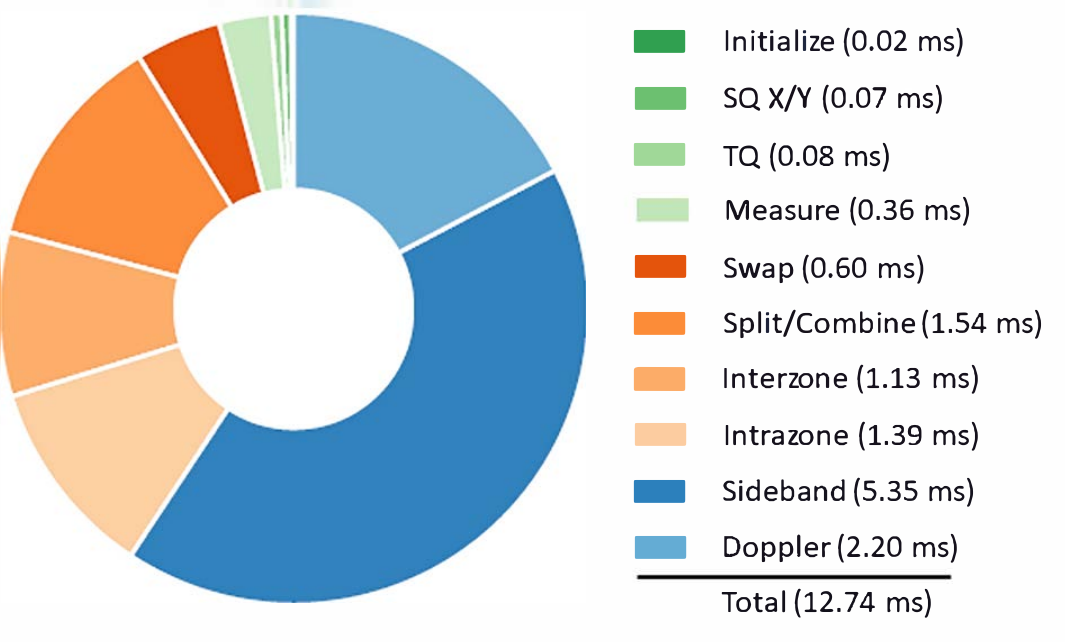}
\caption{The teleported CNOT gate time budget.}
\label{fig:CNOT_time_budget}
\end{center}
\end{figure}

\textbf{Teleported CNOT gate:} The circuit for teleporting a CNOT gate is shown in Fig.~\ref{fig:CNOT_teleportation}a. 
Qubits $q_1$ and $q_2$ are prepared in the state $\frac{1}{\sqrt{2}}\big(\ket{00}+\ket{11}\big)$, then distributed between the two zones through transport operations. Two rounds of CNOT gates (compiled into native $U_{zz}$ and SQ gates) followed by measurements and conditional gates result in a circuit that is logically equivalent to a CNOT controlled on $q_0$ and targeting $q_3$ (see Fig.~\ref{fig:CNOT_teleportation}). A breakdown of the time budget for the circuit is shown in Fig. \ref{fig:CNOT_time_budget}.

Benchmarking the teleported CNOT gate amounts to verifying the
following quantum truth table~\cite{Hofmann2005}:
\begin{equation}
\begin{aligned}[c]
    CNOT:\quad
\end{aligned}
\begin{aligned}[c]
    \ket{00}&\mapsto\ket{00},\quad\ket{++}\mapsto\ket{++},\notag\\
    \ket{01}&\mapsto\ket{11},\quad\ket{+-}\mapsto\ket{+-},\notag\\
    \ket{10}&\mapsto\ket{10},\quad\ket{-+}\mapsto\ket{--},\notag\\
    \ket{11}&\mapsto\ket{01},\quad\ket{--}\mapsto\ket{-+},
\end{aligned}
\end{equation}
where the states are labeled $\ket{q_3\,q_0}$. We prepare $q_0$ and $q_3$ in each state of the $\{\ket{0}, \ket{1}\}$ and $\{\ket{+}, \ket{-}\}$ bases,
apply the circuit in Fig.~\ref{fig:CNOT_teleportation}a and measure in the
appropriate basis. The data is shown in Fig.~\ref{fig:CNOT_teleportation}b.

We define $f_1$ and $f_2$ as the average success probabilities 
over the $\{\ket{0},\ket{1}\}$ and $\{\ket{+}, \ket{-}\}$ 
bases, respectively.
The average fidelity $F_{\rm avg}$ \cite{Nielsen02} of the teleported CNOT gate is bounded according to
\begin{equation}\label{eq: Hofmann bound}
    F_{\rm avg} \ge \frac{4}{5}\big(f_1 + f_2 \big) - \frac{3}{5},
\end{equation}
and does not account SPAM errors, which should be much smaller than the full circuit error.
We find $f_1=0.933(6)$ and $f_2=0.941(5)$, yielding $F_{\rm avg}\ge0.899(6)$.

%\

\textbf{Quantum volume:} The $N=2,3,$ and $4$ tests were performed with four qubit ions loaded into the machine, and the $N=6$ test was performed with six qubit ions loaded into the machine. The $N=2,3,$ and $4$ tests consisted of 100 random circuits repeated 500 times each, and the $N=6$ circuits consisted of 400 circuits repeated 100 times each, all generated by Qiskit~\cite{Qiskit}. All confidence intervals are calculated from the method in Ref.~\cite{Cross19}. The $N=2,3,$ and $4$ QV circuits were run without any optimization or approximations. For example, in the $N=4$ test, each circuit contained exactly 24 TQ gates. For the $N=6$ test we applied an optimization that combines SU(4) gates when the random permutations of qubit pairs calls for gates on the same pair in adjacent steps ~\cite{Cross19}, reducing the average number of TQ gates per circuit from $54$ to $45$. An example of the transport operations for the $N=4$ case is depicted in Fig.~\ref{fig:N=4_Dance}.

\begin{figure}[!t]
\begin{center}
\includegraphics[width=1.0\columnwidth]{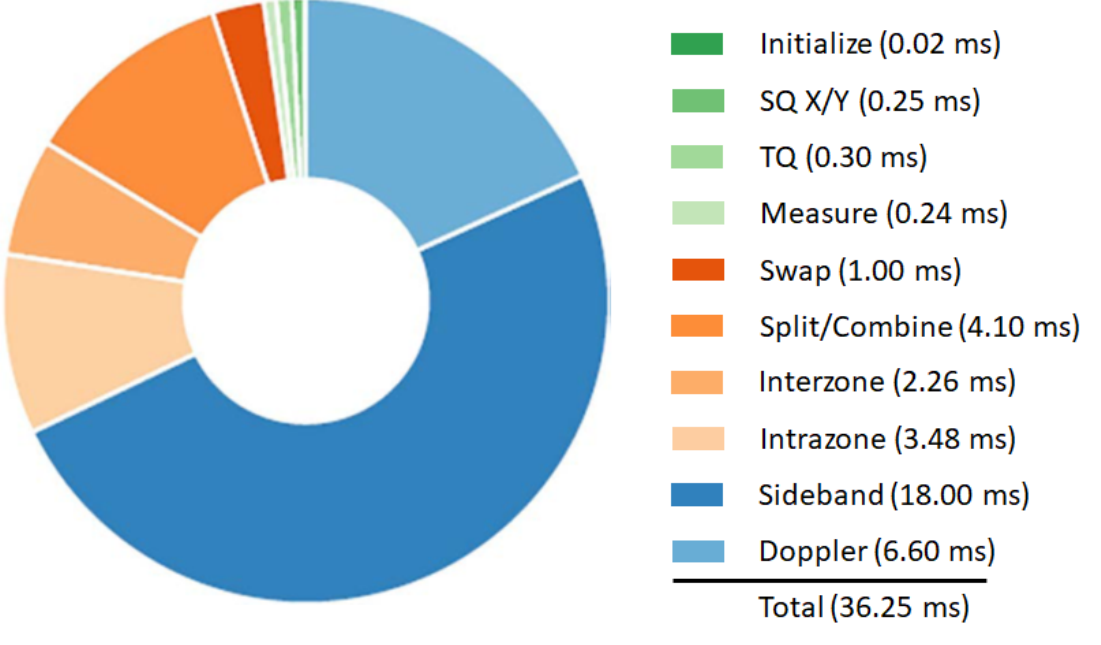}
\caption{The $N=4$ QV test time budget.}
\label{fig:QV_time_budget}
\end{center}
\end{figure}

\begin{figure}[!t]
\begin{center}
\includegraphics[width=1.0\columnwidth]{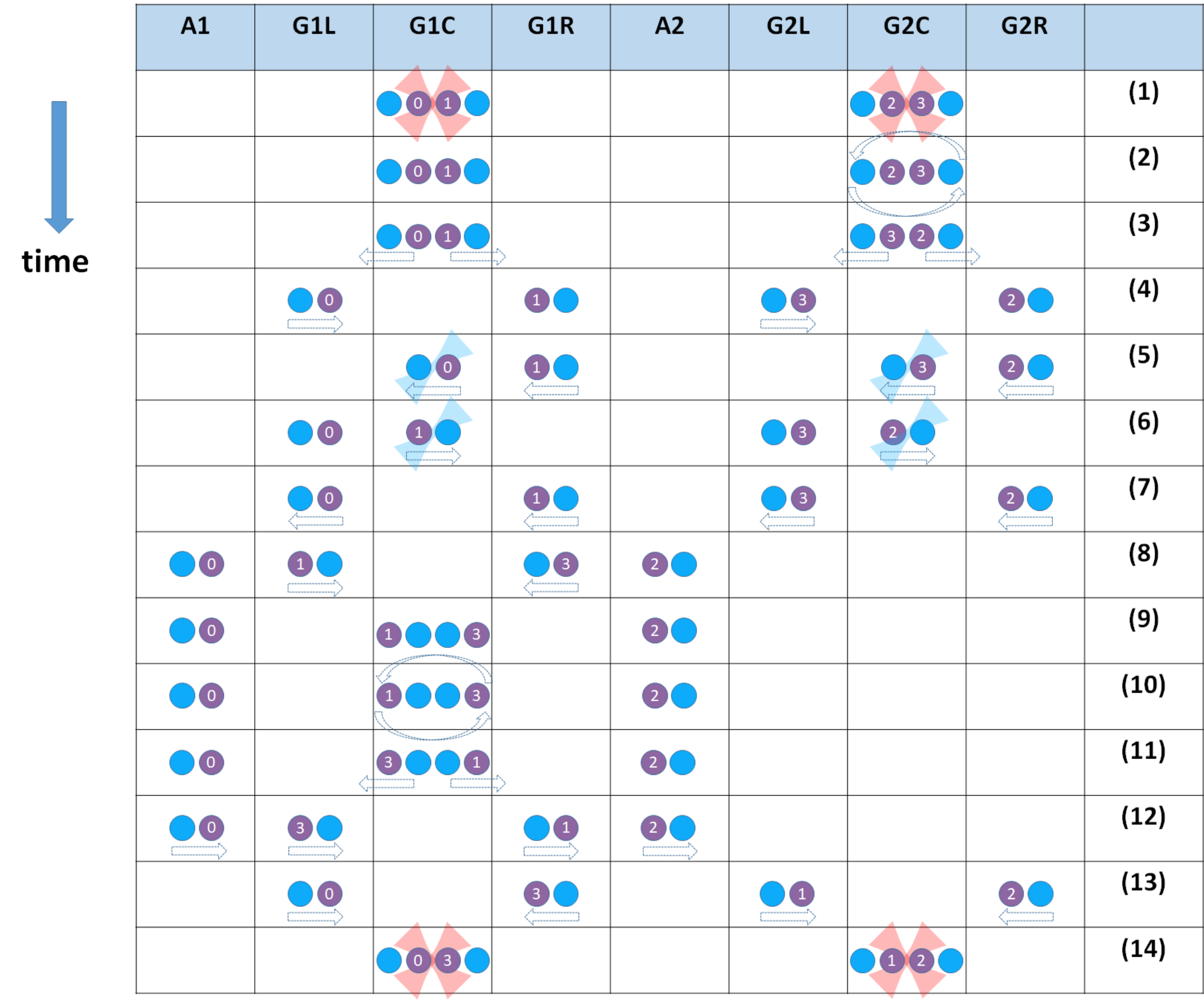}
\caption{An example of the transport operations needed in an $N=4$ QV test circuit.}
\label{fig:N=4_Dance}
\end{center}
\end{figure}

\begin{acknowledgements}
\section{Acknowledgements}
This work was made possible by a large group of people, and the authors would like to thank the entire Honeywell Quantum Solutions team for their many contributions.
\section{Competing Interests}
The authors declare that they have no
competing financial interests.
\section{Correspondence}
 Correspondence and requests for materials
should be addressed to david.hayes@honeywell.com.
\end{acknowledgements}


\begin{thebibliography}{10}
\expandafter\ifx\csname url\endcsname\relax
 \def\url#1{\texttt{#1}}\fi
\expandafter\ifx\csname urlprefix\endcsname\relax\def\urlprefix{URL }\fi
\providecommand{\bibinfo}[2]{#2}
\providecommand{\eprint}[2][]{\url{#2}}

\bibitem{Wineland98}
\bibinfo{author}{Wineland, D.~J.} \emph{et~al.}
\newblock \bibinfo{title}{Experimental issues in coherent quantum-state
 manipulation of trapped atomic ions}.
\newblock \emph{\bibinfo{journal}{Journal of Research of the National Institute
 of Standards and Technology}} \textbf{\bibinfo{volume}{103}},
 \bibinfo{pages}{259--328} (\bibinfo{year}{1998}).

\bibitem{Kielpinski02}
\bibinfo{author}{Kielpinski, D.}, \bibinfo{author}{Monroe, C.} \&
  \bibinfo{author}{Wineland, D.~J.}
\newblock \bibinfo{title}{Architecture for a large-scale ion-trap quantum
  computer}.
\newblock \emph{\bibinfo{journal}{Nature}} \textbf{\bibinfo{volume}{417}},
  \bibinfo{pages}{709--711} (\bibinfo{year}{2002}).
\newblock \urlprefix\url{https://doi.org/10.1038/nature00784}.

\bibitem{Gaebler16}
\bibinfo{author}{Gaebler, J.~P.} \emph{et~al.}
\newblock \bibinfo{title}{High-fidelity universal gate set for
  ${^{9}\mathrm{Be}}^{+}$ ion qubits}.
\newblock \emph{\bibinfo{journal}{Phys. Rev. Lett.}}
  \textbf{\bibinfo{volume}{117}}, \bibinfo{pages}{060505}
  (\bibinfo{year}{2016}).
\newblock
  \urlprefix\url{https://link.aps.org/doi/10.1103/PhysRevLett.117.060505}.

\bibitem{Ballance16}
\bibinfo{author}{Ballance, C.~J.}, \bibinfo{author}{Harty, T.~P.},
  \bibinfo{author}{Linke, N.~M.}, \bibinfo{author}{Sepiol, M.~A.} \&
  \bibinfo{author}{Lucas, D.~M.}
\newblock \bibinfo{title}{High-fidelity quantum logic gates using trapped-ion
  hyperfine qubits}.
\newblock \emph{\bibinfo{journal}{Phys. Rev. Lett.}}
  \textbf{\bibinfo{volume}{117}}, \bibinfo{pages}{060504}
  (\bibinfo{year}{2016}).
\newblock
  \urlprefix\url{https://link.aps.org/doi/10.1103/PhysRevLett.117.060504}.

\bibitem{Christensen19}
\bibinfo{author}{Christensen, J.~E.}, \bibinfo{author}{Hucul, D.},
  \bibinfo{author}{Campbell, W.~C.} \& \bibinfo{author}{Hudson, E.~R.}
\newblock \bibinfo{title}{High fidelity manipulation of a qubit built from a
  synthetic nucleus}  (\bibinfo{year}{2019}).
\newblock \urlprefix\url{https://arxiv.org/pdf/1907.13331.pdf}.
\newblock \eprint{1907.13331}.

\bibitem{Wan19}
\bibinfo{author}{Wan, Y.} \emph{et~al.}
\newblock \bibinfo{title}{Quantum gate teleportation between separated qubits
  in a trapped-ion processor}.
\newblock \emph{\bibinfo{journal}{Science}} \textbf{\bibinfo{volume}{364}},
  \bibinfo{pages}{875--878} (\bibinfo{year}{2019}).
\newblock \urlprefix\url{https://doi.org/10.1126/science.aaw9415}.

\bibitem{Cross19}
\bibinfo{author}{Cross, A.~W.}, \bibinfo{author}{Bishop, L.~S.},
  \bibinfo{author}{Sheldon, S.}, \bibinfo{author}{Nation, P.~D.} \&
  \bibinfo{author}{Gambetta, J.~M.}
\newblock \bibinfo{title}{Validating quantum computers using randomized model
  circuits}.
\newblock \emph{\bibinfo{journal}{Phys. Rev. A}}
  \textbf{\bibinfo{volume}{100}}, \bibinfo{pages}{032328}
  (\bibinfo{year}{2019}).
\newblock \urlprefix\url{https://link.aps.org/doi/10.1103/PhysRevA.100.032328}.

\bibitem{Cirac95}
\bibinfo{author}{Cirac, J.~I.} \& \bibinfo{author}{Zoller, P.}
\newblock \bibinfo{title}{Quantum computations with cold trapped ions}.
\newblock \emph{\bibinfo{journal}{Phys. Rev. Lett.}}
  \textbf{\bibinfo{volume}{74}}, \bibinfo{pages}{4091--4094}
  (\bibinfo{year}{1995}).
\newblock \urlprefix\url{https://link.aps.org/doi/10.1103/PhysRevLett.74.4091}.

\bibitem{Monroe95}
\bibinfo{author}{Monroe, C.}, \bibinfo{author}{Meekhof, D.~M.},
  \bibinfo{author}{King, B.~E.}, \bibinfo{author}{Itano, W.~M.} \&
  \bibinfo{author}{Wineland, D.~J.}
\newblock \bibinfo{title}{Demonstration of a fundamental quantum logic gate}.
\newblock \emph{\bibinfo{journal}{Phys. Rev. Lett.}}
  \textbf{\bibinfo{volume}{75}}, \bibinfo{pages}{4714--4717}
  (\bibinfo{year}{1995}).
\newblock \urlprefix\url{https://link.aps.org/doi/10.1103/PhysRevLett.75.4714}.

\bibitem{Wang17}
\bibinfo{author}{Wang, Y.} \emph{et~al.}
\newblock \bibinfo{title}{Single-qubit quantum memory exceeding ten-minute
  coherence time}.
\newblock \emph{\bibinfo{journal}{Nat. Photonics}}
  \textbf{\bibinfo{volume}{11}}, \bibinfo{pages}{646--650}
  (\bibinfo{year}{2017}).
\newblock \urlprefix\url{https://doi.org/10.1038/s41566-017-0007-1}.

\bibitem{Murali20}
\bibinfo{author}{Murali, P.}, \bibinfo{author}{Debroy, D.~M.},
  \bibinfo{author}{Brown, K.~R.} \& \bibinfo{author}{Martonosi, M.}
\newblock \bibinfo{title}{Architecting noisy intermediate-scale trapped ion
  quantum computers}  (\bibinfo{year}{2020}).
\newblock \urlprefix\url{https://arxiv.org/pdf/2004.04706.pdf}.
\newblock \eprint{2004.04706}.

\bibitem{Monroe14}
\bibinfo{author}{Monroe, C.} \emph{et~al.}
\newblock \bibinfo{title}{Large-scale modular quantum-computer architecture
  with atomic memory and photonic interconnects}.
\newblock \emph{\bibinfo{journal}{Phys. Rev. A}} \textbf{\bibinfo{volume}{89}},
  \bibinfo{pages}{022317} (\bibinfo{year}{2014}).
\newblock \urlprefix\url{https://link.aps.org/doi/10.1103/PhysRevA.89.022317}.

\bibitem{Hucul15}
\bibinfo{author}{Hucul, D.} \emph{et~al.}
\newblock \bibinfo{title}{Modular entanglement of atomic qubits using photons
  and phonons}.
\newblock \emph{\bibinfo{journal}{Nat. Phys.}} \textbf{\bibinfo{volume}{11}},
  \bibinfo{pages}{37--42} (\bibinfo{year}{2015}).
\newblock \urlprefix\url{https://doi.org/10.1038/nphys3150DO}.

\bibitem{Home09a}
\bibinfo{author}{Home, J.~P.} \emph{et~al.}
\newblock \bibinfo{title}{Complete methods set for scalable ion trap quantum
  information processing}.
\newblock \emph{\bibinfo{journal}{Science}} \textbf{\bibinfo{volume}{325}},
  \bibinfo{pages}{1227--1230} (\bibinfo{year}{2009}).
\newblock \urlprefix\url{https://science.sciencemag.org/content/325/5945/1227}.
\newblock
  \eprint{https://science.sciencemag.org/content/325/5945/1227.full.pdf}.

\bibitem{Kaufmann17}
\bibinfo{author}{Kaufmann, H.} \emph{et~al.}
\newblock \bibinfo{title}{Scalable creation of long-lived multipartite
  entanglement}.
\newblock \emph{\bibinfo{journal}{Phys. Rev. Lett.}}
  \textbf{\bibinfo{volume}{119}}, \bibinfo{pages}{150503}
  (\bibinfo{year}{2017}).
\newblock
  \urlprefix\url{https://link.aps.org/doi/10.1103/PhysRevLett.119.150503}.

\bibitem{Lekitsche17}
\bibinfo{author}{Lekitsch, B.} \emph{et~al.}
\newblock \bibinfo{title}{Blueprint for a microwave trapped ion quantum
  computer}.
\newblock \emph{\bibinfo{journal}{Science Advances}}
  \textbf{\bibinfo{volume}{3}} (\bibinfo{year}{2017}).
\newblock \urlprefix\url{https://advances.sciencemag.org/content/3/2/e1601540}.
\newblock
  \eprint{https://advances.sciencemag.org/content/3/2/e1601540.full.pdf}.

\bibitem{LabaziewiczThesis08}
\bibinfo{author}{Labaziewicz, J.}
\newblock \emph{\bibinfo{title}{High Fidelity Quantum Gates with Ions in
  Cryogenic Microfabricated Ion Traps}}.
\newblock Ph.D. thesis, \bibinfo{school}{MIT} (\bibinfo{year}{2008}).
\newblock
  \urlprefix\url{http://web.mit.edu/~cua/www/quanta/LabaziewiczThesis.pdf}.

\bibitem{Maunz16}
\bibinfo{author}{Maunz, P. L.~W.}
\newblock \bibinfo{title}{High optical access trap 2.0.}
\newblock \emph{\bibinfo{journal}{SAND2016-0796R}}  (\bibinfo{year}{2016}).
\newblock
  \urlprefix\url{https://prod-ng.sandia.gov/techlib-noauth/access-control.cgi/2016/160796r.pdf}.

\bibitem{Bowler12}
\bibinfo{author}{Bowler, R.} \emph{et~al.}
\newblock \bibinfo{title}{Coherent diabatic ion transport and separation in a
  multizone trap array}.
\newblock \emph{\bibinfo{journal}{Phys. Rev. Lett.}}
  \textbf{\bibinfo{volume}{109}}, \bibinfo{pages}{080502}
  (\bibinfo{year}{2012}).
\newblock
  \urlprefix\url{https://link.aps.org/doi/10.1103/PhysRevLett.109.080502}.

\bibitem{Kaushal20}
\bibinfo{author}{Kaushal, V.} \emph{et~al.}
\newblock \bibinfo{title}{Shuttling-based trapped-ion quantum information
  processing}.
\newblock \emph{\bibinfo{journal}{AVS Quantum Science}}
  \textbf{\bibinfo{volume}{2}}, \bibinfo{pages}{014101} (\bibinfo{year}{2020}).
\newblock \urlprefix\url{https://doi.org/10.1116/1.5126186}.

\bibitem{Barrett03}
\bibinfo{author}{Barrett, M.~D.} \emph{et~al.}
\newblock \bibinfo{title}{Sympathetic cooling of ${}^{9}{\mathrm{be}}^{+}$ and
  ${}^{24}{\mathrm{mg}}^{+}$ for quantum logic}.
\newblock \emph{\bibinfo{journal}{Phys. Rev. A}} \textbf{\bibinfo{volume}{68}},
  \bibinfo{pages}{042302} (\bibinfo{year}{2003}).
\newblock \urlprefix\url{https://link.aps.org/doi/10.1103/PhysRevA.68.042302}.

\bibitem{Home09}
\bibinfo{author}{Home, J.~P.} \emph{et~al.}
\newblock \bibinfo{title}{Memory coherence of a sympathetically cooled
  trapped-ion qubit}.
\newblock \emph{\bibinfo{journal}{Phys. Rev. A}} \textbf{\bibinfo{volume}{79}},
  \bibinfo{pages}{050305} (\bibinfo{year}{2009}).
\newblock \urlprefix\url{https://link.aps.org/doi/10.1103/PhysRevA.79.050305}.

\bibitem{Nielsen00}
\bibinfo{author}{Nielsen, M.~A.} \& \bibinfo{author}{Chuang, I.~L.}
\newblock \emph{\bibinfo{title}{Quantum Computation and Quantum Information}}
  (\bibinfo{publisher}{Cambridge University Press}, \bibinfo{year}{2000}).

\bibitem{Palmero14}
\bibinfo{author}{Palmero, M.}, \bibinfo{author}{Bowler, R.},
  \bibinfo{author}{Gaebler, J.~P.}, \bibinfo{author}{Leibfried, D.} \&
  \bibinfo{author}{Muga, J.~G.}
\newblock \bibinfo{title}{Fast transport of mixed-species ion chains within a
  paul trap}.
\newblock \emph{\bibinfo{journal}{Phys. Rev. A}} \textbf{\bibinfo{volume}{90}},
  \bibinfo{pages}{053408} (\bibinfo{year}{2014}).
\newblock \urlprefix\url{https://link.aps.org/doi/10.1103/PhysRevA.90.053408}.

\bibitem{Home06}
\bibinfo{author}{Home, J.~P.} \& \bibinfo{author}{Steane, A.~M.}
\newblock \bibinfo{title}{Electrode configurations for fast separation of
  trapped ions}.
\newblock \emph{\bibinfo{journal}{Quantum Information and Computation}}
  \textbf{\bibinfo{volume}{6}}, \bibinfo{pages}{289--325}
  (\bibinfo{year}{2006}).
\newblock \urlprefix\url{https://dl.acm.org/doi/10.5555/2012086.2012087}.

\bibitem{Splatt09}
\bibinfo{author}{Splatt, F.} \emph{et~al.}
\newblock \bibinfo{title}{Deterministic reordering of $^{40}$ca$^+$ ions in a
  linear segmented paul trap}.
\newblock \emph{\bibinfo{journal}{New Journal of Phyics}}
  \textbf{\bibinfo{volume}{11}}, \bibinfo{pages}{103008}
  (\bibinfo{year}{2009}).
\newblock \urlprefix\url{https://doi.org/10.1088/1367-2630/11/10/103008}.

\bibitem{Haberman1979}
\bibinfo{author}{Haberman, N.}
\newblock \bibinfo{title}{Parallel neighbor sort (or the glory of the induction
  principle)}.
\newblock \emph{\bibinfo{journal}{CMU Computer Science Report}}
  (\bibinfo{year}{1979}).

\bibitem{Molmer00}
\bibinfo{author}{S\o{}rensen, A.} \& \bibinfo{author}{M\o{}lmer, K.}
\newblock \bibinfo{title}{Entanglement and quantum computation with ions in
  thermal motion}.
\newblock \emph{\bibinfo{journal}{Phys. Rev. A}} \textbf{\bibinfo{volume}{62}},
  \bibinfo{pages}{022311} (\bibinfo{year}{2000}).
\newblock \urlprefix\url{https://link.aps.org/doi/10.1103/PhysRevA.62.022311}.

\bibitem{Lee05}
\bibinfo{author}{Lee, P.~J.} \emph{et~al.}
\newblock \bibinfo{title}{Phase control of trapped ion quantum gates}.
\newblock \emph{\bibinfo{journal}{Journal of Optics B: Quantum and
  Semiclassical Optics}} \textbf{\bibinfo{volume}{7}},
  \bibinfo{pages}{S371--S383} (\bibinfo{year}{2005}).
\newblock \urlprefix\url{https://doi.org/10.1088/1464-4266/7/10/025}.

\bibitem{Baldwin19}
\bibinfo{author}{Baldwin, C.~H.}, \bibinfo{author}{Bjork, B.~J.},
  \bibinfo{author}{Gaebler, J.~P.}, \bibinfo{author}{Hayes, D.} \&
  \bibinfo{author}{Stack, D.}
\newblock \bibinfo{title}{Subspace benchmarking high-fidelity entangling
  operations with trapped ions}.
\newblock \emph{\bibinfo{journal}{Phys. Rev. Research}}
  \textbf{\bibinfo{volume}{2}}, \bibinfo{pages}{013317} (\bibinfo{year}{2020}).
\newblock
  \urlprefix\url{https://link.aps.org/doi/10.1103/PhysRevResearch.2.013317}.

\bibitem{Viola99}
\bibinfo{author}{Viola, L.}, \bibinfo{author}{Knill, E.} \&
  \bibinfo{author}{Lloyd, S.}
\newblock \bibinfo{title}{Dynamical decoupling of open quantum systems}.
\newblock \emph{\bibinfo{journal}{Phys. Rev. Lett.}}
  \textbf{\bibinfo{volume}{82}}, \bibinfo{pages}{2417--2421}
  (\bibinfo{year}{1999}).
\newblock \urlprefix\url{https://link.aps.org/doi/10.1103/PhysRevLett.82.2417}.

\bibitem{Olmschenk07}
\bibinfo{author}{Olmschenk, S.} \emph{et~al.}
\newblock \bibinfo{title}{Manipulation and detection of a trapped yb+ hyperfine
  qubit}.
\newblock \emph{\bibinfo{journal}{Phys. Rev. A}} \textbf{\bibinfo{volume}{76}},
  \bibinfo{pages}{052314} (\bibinfo{year}{2007}).
\newblock \urlprefix\url{https://link.aps.org/doi/10.1103/PhysRevA.76.052314}.

\bibitem{Magesan11}
\bibinfo{author}{Magesan, E.}, \bibinfo{author}{Gambetta, J.~M.} \&
  \bibinfo{author}{Emerson, J.}
\newblock \bibinfo{title}{Scalable and robust randomized benchmarking of
  quantum processes}.
\newblock \emph{\bibinfo{journal}{Phys. Rev. Lett.}}
  \textbf{\bibinfo{volume}{106}}, \bibinfo{pages}{180504}
  (\bibinfo{year}{2011}).
\newblock
  \urlprefix\url{https://link.aps.org/doi/10.1103/PhysRevLett.106.180504}.

\bibitem{MonroeCooling95}
\bibinfo{author}{Monroe, C.} \emph{et~al.}
\newblock \bibinfo{title}{Resolved-sideband raman cooling of a bound atom to
  the 3d zero-point energy}.
\newblock \emph{\bibinfo{journal}{Phys. Rev. Lett.}}
  \textbf{\bibinfo{volume}{75}}, \bibinfo{pages}{4011--4014}
  (\bibinfo{year}{1995}).
\newblock \urlprefix\url{https://link.aps.org/doi/10.1103/PhysRevLett.75.4011}.

\bibitem{Jordan2019}
\bibinfo{author}{Jordan, E.} \emph{et~al.}
\newblock \bibinfo{title}{Near ground-state cooling of two-dimensional
  trapped-ion crystals with more than 100 ions}.
\newblock \emph{\bibinfo{journal}{Phys. Rev. Lett.}}
  \textbf{\bibinfo{volume}{122}}, \bibinfo{pages}{053603}
  (\bibinfo{year}{2019}).
\newblock
  \urlprefix\url{https://link.aps.org/doi/10.1103/PhysRevLett.122.053603}.

\bibitem{Gambetta12}
\bibinfo{author}{Gambetta, J.~M.} \emph{et~al.}
\newblock \bibinfo{title}{Characterization of addressability by simultaneous
  randomized benchmarking}.
\newblock \emph{\bibinfo{journal}{Phys. Rev. Lett.}}
  \textbf{\bibinfo{volume}{109}}, \bibinfo{pages}{240504}
  (\bibinfo{year}{2012}).
\newblock
  \urlprefix\url{https://link.aps.org/doi/10.1103/PhysRevLett.109.240504}.

\bibitem{Barrett04}
\bibinfo{author}{Barrett, M.} \emph{et~al.}
\newblock \bibinfo{title}{Deterministic quantum teleportation of atomic
  qubits}.
\newblock \emph{\bibinfo{journal}{Nature}} \textbf{\bibinfo{volume}{429}},
  \bibinfo{pages}{737} (\bibinfo{year}{2004}).
\newblock \urlprefix\url{https://doi.org/10.1038/nature02608}.

\bibitem{Negnevitsky18}
\bibinfo{author}{Negnevitsky, V.} \emph{et~al.}
\newblock \bibinfo{title}{Repeated multi-qubit readout and feedback with a
  mixed species trapped-ion register}.
\newblock \emph{\bibinfo{journal}{Nature}} \textbf{\bibinfo{volume}{429}},
  \bibinfo{pages}{737} (\bibinfo{year}{2004}).
\newblock \urlprefix\url{https://doi.org/10.1038/nature02608}.

\bibitem{Plenio2000}
\bibinfo{author}{Eisert, J.}, \bibinfo{author}{Jacobs, K.},
  \bibinfo{author}{Papadopoulos, P.} \& \bibinfo{author}{Plenio, M.~B.}
\newblock \bibinfo{title}{Optimal local implementation of nonlocal quantum
  gates}.
\newblock \emph{\bibinfo{journal}{Phys. Rev. A}} \textbf{\bibinfo{volume}{62}},
  \bibinfo{pages}{052317} (\bibinfo{year}{2000}).
\newblock \urlprefix\url{https://link.aps.org/doi/10.1103/PhysRevA.62.052317}.

\bibitem{McClean16}
\bibinfo{author}{McClean, J.~R.}, \bibinfo{author}{Romero, J.},
  \bibinfo{author}{Babbush, R.} \& \bibinfo{author}{Aspuru-Guzik, A.}
\newblock \bibinfo{title}{The theory of variational hybrid quantum-classical
  algorithms}.
\newblock \emph{\bibinfo{journal}{New J. Phys.}} \textbf{\bibinfo{volume}{18}},
  \bibinfo{pages}{023023} (\bibinfo{year}{2016}).
\newblock \urlprefix\url{https://doi.org/10.1088/1367-2630/18/2/023023}.

\bibitem{Farhi14}
\bibinfo{author}{Farhi, E.} \& \bibinfo{author}{Goldstone, J.}
\newblock \bibinfo{title}{A quantum approximate optimization algorithm}
  (\bibinfo{year}{2014}).
\newblock \urlprefix\url{https://arxiv.org/pdf/1411.4028.pdf}.
\newblock \eprint{1411.4028}.

\bibitem{Aaronson16}
\bibinfo{author}{Aaronson, S.} \& \bibinfo{author}{Chen, L.}
\newblock \bibinfo{title}{Complexity-theoretic foundations of quantum supremacy
  experiments}  (\bibinfo{year}{2016}).
\newblock \urlprefix\url{https://arxiv.org/pdf/1612.05903.pdf}.
\newblock \eprint{1612.05903}.

\bibitem{Qiskit}
\bibinfo{author}{Abraham, H.} \& \bibinfo{author}{\textit{et al.}}
\newblock \bibinfo{title}{Qiskit: An open-source framework for quantum
  computing} (\bibinfo{year}{2019}).

\bibitem{Jurcevic20}
\bibinfo{author}{Jucevic, P.} \& \bibinfo{author}{\textit{et al.}}
\newblock \bibinfo{title}{Demonstration of quantum volume 64 on a
  superconducting quantum computing system}  (\bibinfo{year}{2020}).
\newblock \urlprefix\url{https://arxiv.org/abs/2008.08571}.
\newblock \eprint{2008.08571}.

\bibitem{Mourik2020}
\bibinfo{author}{Mourik, M. W.~v.} \emph{et~al.}
\newblock \bibinfo{title}{Coherent rotations of qubits within a multi-species
  ion-trap quantum computer}  (\bibinfo{year}{2020}).
\newblock \urlprefix\url{https://arxiv.org/pdf/2001.02440.pdf}.
\newblock \eprint{2001.02440}.

\bibitem{Mount_2013}
\bibinfo{author}{Mount, E.} \emph{et~al.}
\newblock \bibinfo{title}{Single qubit manipulation in a microfabricated
  surface electrode ion trap}.
\newblock \emph{\bibinfo{journal}{New Journal of Physics}}
  \textbf{\bibinfo{volume}{15}}, \bibinfo{pages}{093018}
  (\bibinfo{year}{2013}).
\newblock \urlprefix\url{https://doi.org/10.1088/1367-2630/15/9/093018}.

\bibitem{Mehta2020}
\bibinfo{author}{Mehta, K.~K.} \emph{et~al.}
\newblock \bibinfo{title}{Integrated optical multi-ion quantum logic}
  (\bibinfo{year}{2020}).
\newblock \urlprefix\url{https://arxiv.org/pdf/2002.02258.pdf}.
\newblock \eprint{2002.02258}.

\bibitem{Kovalev2013}
\bibinfo{author}{Kovalev, A.~A.} \& \bibinfo{author}{Pryadko, L.~P.}
\newblock \bibinfo{title}{Quantum kronecker sum-product low-density
  parity-check codes with finite rate}.
\newblock \emph{\bibinfo{journal}{Phys. Rev. A}} \textbf{\bibinfo{volume}{88}},
  \bibinfo{pages}{012311} (\bibinfo{year}{2013}).
\newblock \urlprefix\url{https://link.aps.org/doi/10.1103/PhysRevA.88.012311}.

\bibitem{Labaziewicz08}
\bibinfo{author}{Labaziewicz, J.} \emph{et~al.}
\newblock \bibinfo{title}{Temperature dependence of electric field noise above
  gold surfaces}.
\newblock \emph{\bibinfo{journal}{Phys. Rev. Lett.}}
  \textbf{\bibinfo{volume}{101}}, \bibinfo{pages}{180602}
  (\bibinfo{year}{2008}).
\newblock
  \urlprefix\url{https://link.aps.org/doi/10.1103/PhysRevLett.101.180602}.

\bibitem{Blakestad10}
\bibinfo{author}{Blakestad, R.~B.}
\newblock \emph{\bibinfo{title}{Transport of Trapped-ion Qubits within a
  Scalable Quantum Processor}}.
\newblock Ph.D. thesis, \bibinfo{school}{University of Colorado}
  (\bibinfo{year}{2010}).
\newblock
  \urlprefix\url{https://www.nist.gov/system/files/documents/2017/05/09/blakestad2010thesis.pdf}.

\bibitem{Biercuk_2011}
\bibinfo{author}{Biercuk, M.~J.}, \bibinfo{author}{Doherty, A.~C.} \&
  \bibinfo{author}{Uys, H.}
\newblock \bibinfo{title}{Dynamical decoupling sequence construction as a
  filter-design problem}.
\newblock \emph{\bibinfo{journal}{Journal of Physics B: Atomic, Molecular and
  Optical Physics}} \textbf{\bibinfo{volume}{44}}, \bibinfo{pages}{154002}
  (\bibinfo{year}{2011}).
\newblock \urlprefix\url{https://doi.org/10.1088/0953-4075/44/15/154002}.

\bibitem{Harper19}
\bibinfo{author}{Harper, R.}, \bibinfo{author}{Flammia, S.~T.} \&
  \bibinfo{author}{Wallman, J.~J.}
\newblock \bibinfo{title}{Efficient learning of quantum noise}
  (\bibinfo{year}{2019}).
\newblock \urlprefix\url{https://arxiv.org/pdf/1907.13022.pdf}.
\newblock \eprint{1907.13022}.

\bibitem{Meier06}
\bibinfo{author}{Meier, A.~M.}
\newblock \emph{\bibinfo{title}{Randomized Benchmarking of Clifford
  Operators}}.
\newblock Ph.D. thesis, \bibinfo{school}{University of Colorado}
  (\bibinfo{year}{2006}).
\newblock \urlprefix\url{https://arxiv.org/abs/1811.10040}.
\newblock \eprint{1811.10040}.

\bibitem{Hofmann2005}
\bibinfo{author}{Hofmann, H.~F.}
\newblock \bibinfo{title}{Complementary classical fidelities as an efficient
  criterion for the evaluation of experimentally realized quantum operations}.
\newblock \emph{\bibinfo{journal}{Phys. Rev. Lett.}}
  \textbf{\bibinfo{volume}{94}}, \bibinfo{pages}{160504}
  (\bibinfo{year}{2005}).
\newblock
  \urlprefix\url{https://link.aps.org/doi/10.1103/PhysRevLett.94.160504}.

\bibitem{Nielsen02}
\bibinfo{author}{Nielsen, M.~A.}
\newblock \bibinfo{title}{A simple formula for the average gate fidelity of a
  quantum dynamical operation}.
\newblock \emph{\bibinfo{journal}{Physics Letters A}}
  \textbf{\bibinfo{volume}{303}}, \bibinfo{pages}{249--252}
  (\bibinfo{year}{2002}).
\newblock \urlprefix\url{https://doi.org/10.1016/S0375-9601(02)01272-0}.

\end{thebibliography}
\end{document}